\documentclass[useAMS,usenatbib,usegraphicx,epsfig]{aa}
\usepackage{txfonts,graphicx,epsfig}

\def\eps@scaling{.95}
\def\epsscale#1{\gdef\eps@scaling{#1}}
\def\plotone#1{\centering \leavevmode
\epsfxsize=\eps@scaling\columnwidth \epsfbox{#1}}

\def\micron{\mu {\rm m}}
\begin{document}

\title{How to identify the youngest protostars}

\titlerunning{How to identify the youngest protostars}

\author{D.~Stamatellos, A.~P.~Whitworth, D.~F.~A.~Boyd, \& S.~P.~Goodwin}

\authorrunning{D.~Stamatellos, A.~P.~Whitworth, et al.}

\offprints{D.~Stamatellos\\ \email{D.Stamatellos@astro.cf.ac.uk}}

\institute{School of Physics \& Astronomy, Cardiff University, 
        5 The Parade, Cardiff CF24 3YB, Wales, UK}

\date{Received ..., 2005; accepted ... , 2005}

\abstract{
 We study the transition from a prestellar core to a Class 0 protostar, 
using SPH to simulate the dynamical evolution, and a Monte Carlo radiative 
transfer code to generate the SED and isophotal maps. For a prestellar core 
illuminated by the standard interstellar radiation field, the luminosity 
is low and the SED peaks at $\sim\!190\,\micron$. Once a protostar has 
formed, the luminosity rises (due to a growing contribution from accretion 
onto the protostar) and the peak of the SED shifts to shorter wavelengths 
($80\;{\rm to}\,100\,\micron$). However, by the end of the Class 0 phase, 
the accretion rate is falling, the luminosity has decreased, and the peak of 
the SED shifts back towards longer wavelengths ($90\;{\rm to}\,150\,\micron$). 
In our simulations, the density of material around the protostar remains 
sufficiently high well into the Class 0 phase that the protostar only 
becomes visible in the NIR if it is displaced from the centre dynamically. 
Raw submm/mm maps of Class 0 protostars tend to be much more centrally 
condensed than those of prestellar cores. However, when convolved with a 
typical telescope beam, the difference in central concentration is less 
marked, although the Class 0 protostars appear more circular. Our results 
suggest that, if a core is deemed to be prestellar on the basis of having 
no associated IRAS source, no cm radio emission, and no 
outflow, but it has a circular appearance and an SED which peaks at 
wavelengths below $\sim 170\,\micron$, it may well contain a very 
young Class 0 protostar.

 }

\maketitle


\section{Introduction} 

The details of how molecular clouds form, why they collapse, and how 
the collapse proceeds to form stars are not very well understood. Over 
the last two decades, observations of the different stages of this 
process have lead to an evolutionary scenario of star formation: 
starless core $\rightarrow$ prestellar core $\rightarrow$ class 0 
object $\rightarrow$ class I object $\rightarrow $ class II object 
$\rightarrow$ class III object (see Andr\'{e} et al. 2000).

Starless cores (e.g. Myers et al. 1983; Myers \& Benson 1983) are 
dense cores in molecular clouds in which there is no evidence that 
star formation has occurred (i.e. no IRAS detection, Beichman et al. 
1986). Some of these starless cores are thought to be close to 
collapse or already collapsing, and they are labelled prestellar 
cores (Ward-Thompson et al. 1994, 1999); for the rest, it is not 
always clear whether they are in hydrostatic equilibrium (e.g. 
Alves et al. 2001  or transient structures 
within a turbulent cloud (e.g. Ballesteros-Paredes et al. 2003). 
Prestellar cores have been observed both in 
relative isolation (e.g. L1544, L63; Ward-Thompson et al. 1999) and 
in protoclusters (e.g. $\rho$~Oph, Motte et al.~1998; NGC 2068/2071, 
Motte et al.~2001). They have typical sizes $(1-20)\times10^3$~AU 
and masses from $0.05$ to $35\,{\rm M}_{\sun}$.  From statistical arguments 
(e.g. Andr\'e et al. 2000) it is inferred that prestellar cores live 
only a few million years.

Class 0 objects represent the stage when a protostar has just been 
formed in the centre of the core, but the protostar is still less 
massive than its surrounding envelope (Andr\'e et al. 1993). The 
protostar is deeply embedded in the core and cannot be observed 
directly, but its presence can often be inferred from bipolar 
molecular outflows or compact centimetre radio emission. The Class 
0 phase last for a few $10^4\,{\rm years}$ and is characterised by a 
large accretion rate ($\ga 10^{-5}\,{\rm M}_{\sun}$/yr). Most of 
the mass of the protostar is delivered during this phase (e.g. 
Whitworth \& Ward-Thompson 2001).

Once the protostar becomes more massive than its surrounding envelope, 
the core enters the Class I phase. Accretion onto the central protostar, 
or onto the disc around the protostar, continues, but at a lower rate 
($\la 10^{-6}\,{\rm M}_{\sun}$/yr). This phase lasts for a few $10^5\,
{\rm years}$, and delivers most of the rest of the mass of the protostar.

Finally, Class II and Class III objects correspond to even later stages 
of star formation, when most of the envelope has disappeared, and the 
star is visible in the optical. Class I objects are classical T Tauri 
stars (CTTSs), with ongoing accretion (but at much lower levels than 
before, $\la 10^{-8}\,{\rm M}_{\sun}$/yr) and well defined discs. Class 
III objects are weak-line T Tauri stars (WTTs) with little sign of 
accretion  and no inner disc.

In their quest to identify the youngest protostars, Andr\'e et al. 
(1993, 2000) set 3 criteria for classifying an object as Class 0: (i) 
the presence of a central luminosity source, as indicated by the 
detection of a compact centimetre radio source, or a bipolar molecular 
outflow, or internal heating; (ii) centrally peaked but extended 
submillimetre continuum emission, indicating the presence of an 
envelope around the central source; and (iii) a high ratio ($>0.005$) 
of submillimetre to bolometric luminosity, where the submillimetre is 
defined as $\lambda \ga 350\,\micron$. Criteria (ii) and (iii) are 
used to distinguish between Class 0 and Class I objects, whereas 
criterion (i) is used to distinguish between Class 0 objects and 
prestellar cores.  However, criterion (i) may be inadequate to 
distinguish between the youngest Class 0 objects and the oldest 
prestellar cores: (a) the cm emission may be be too weak to be 
observed, especially in the earliest stages of protostar formation; 
(b) the region may be too complex for the molecular outflows to be 
detectable; and (c) internal heating may be difficult to establish, 
if the protostar is deeply embedded in the core.

The origin of radio emission from Class 0 objects is uncertain (cf. Gibb 
1999). Theoretical models show that once a protostar forms at the centre 
of a collapsing core, the infalling gas accelerates to supersonic velocities 
and an accretion shock develops on the surface of the protostar and/or on 
the surface of the disc that surrounds the protostar (Winkler \& Newman 
1980, Cassen \& Moosman 1981). The heating provided by the shock ionises 
the surrounding gas, which then emits free-free radio radiation (e.g. 
Bertout 1983; Neufeld \& Hollenbach 1994, 1996). The radio luminosity 
depends on the mass of the protostar, the accretion rate and the observer's 
viewing angle. For a low-mass protostar, $\la {\rm M}_{\sun}$, the 
predicted flux is below detection limits  (Neufeld \& Hollenbach 1996), 
unless the accretion rate is high, $\ga 10^{-5}\,{\rm M}_{\sun}$/yr. Another 
possibility is that the radio emission is produced by an ionised disc wind 
(e.g. Martin 1996), or from a partially ionised jet that emanates from the 
protostar and propagates into the collapsing envelope producing shocks 
(Curiel et al. 1987). The detection of radio emission from many Class 0 
protostars, using the VLA (e.g. Bontemps et al. 1995), points toward the 
latter two explanations.

Recently Young et al. (2004), using the Spitzer space telescope, detected 
NIR radiation from L1014, a dense core that was previously classified as 
prestellar.  The detection of NIR radiation strongly suggests that this is 
a young Class 0 object, and supports the view that even younger Class 0 
objects may not be detectable in the NIR, with IRAS, or even with Spitzer. 

The goal of this paper is to study the transition from prestellar cores to 
Class 0 objects, and to seek new criteria for distinguishing between these 
two stages, in particular criteria which can be used before the standard 
signatures of protostar birth, such as compact radio emission and/or bipolar 
molecular outflows, become detectable. We use SPH to simulate to the dynamics 
of a collapsing molecular core, and a Monte Carlo code to treat the transfer 
of radiation within the core. The radiative transfer is treated fully in 3D, 
which is important for the correct interpretation of the observations (cf. 
Boss \& Yorke 1990; Whitney et al. 2003a, 2003b; Steinacker et al. 2004). We 
use a newly developed method for performing Monte Carlo radiative transfer 
simulations on SPH density fields, which constructs radiative transfer cells 
using the SPH tree structure (Stamatellos 2003; Stamatellos \& Whitworth 2005, 
hereafter Paper I).

In Section 2 we describe the SPH simulation of the collapse of a turbulent 
molecular core. In Section 3 we describe the radiative transfer method, 
focusing on how we construct radiative transfer cells. We also discuss the 
radiation sources and the properties of the dust used in our model. In 
Section 4 we present our results for the dust temperature fields, SEDs and 
isophotal maps of prestellar cores and young Class 0 protostars. In Section 
5 we discuss how SEDs and isophotal maps at submm and mm wavelengths might 
be used to distinguish between late-phase prestellar cores and early-phase 
Class 0 objects.  In Section 6 we summarise our results.

\section{The SPH simulation} \label{S:SPH}

We use the Smoothed Particle Hydrodynamics code {\sc dragon} (Goodwin et 
al. 2004a) to simulate the collapse of a turbulent molecular core and the 
resulting star formation.  Here, we briefly describe the main elements 
of the model. For a more detailed discussion we refer to Goodwin et al. 
(2004b).

The initial conditions in the core, before the start of collapse, are 
dictated by observations of prestellar cores (e.g. Ward-Thompson et al. 
1999, Alves et al. 2001). Prestellar cores appear to have approximately 
uniform density in their central regions, and the density then falls off 
in the envelope. If the density in the envelope is fitted with a power 
law, $n(r) \propto r^{-\eta}\,$, then $\eta \sim 2\,{\rm to}\,4 \,$. Here 
$\eta \sim 2$ is characteristic of more extended prestellar cores in 
dispersed star formation regions (e.g. L1544, L63 and L43), whereas $\eta 
\sim 4$ is characteristic of more compact cores in protoclusters (e.g. 
$\rho$ Oph and NGC2068/2071).  These features can be represented by a 
Plummer-like density profile (Plummer 1915), 
\begin{equation} 
\rho=\frac{\rho_{_0}}{\left(1+(r/r_{_0})^2\right)^2}\,,\;\;\;\;r < r_{_{\rm B}}
\end{equation}
The density profile is almost flat for $r < r_{_0}$, and falls off as 
$r^{-4}$ in the outer envelope ($r > r_{_0}$). We set $\rho_{_0} = 3 
\times 10^{-18}\,{\rm g\,cm}^{-3}$, $r_{_0}=5,000\,{\rm AU}$, and 
$r_{_{\rm B}} = 50,000\,{\rm AU}$. The core mass is then 5.4 M$_{\sun}$. 
The core is initially isothermal at 10~K, hence the thermal virial ratio 
is $\alpha_{_{\rm THERM}} \equiv E_{_{\rm THERM}} / |E_{_{\rm GRAV}}| = 
0.45$. Observations of star forming cores (e.g. Jijina et al. 1999) show 
that their line widths have a non-thermal contribution, which may be due 
to turbulence. We impose a low level of turbulence ($\alpha_{_{\rm TURB}} 
\equiv E_{_{\rm TURB}} / |E_{_{\rm GRAV}}| = 0.05$) with a power spectrum 
$P(k) \propto k^{-4}$ so as to mimic the observed scaling relation between 
linear size and internal velocity dispersion (Larson 1981).

Collapse of the turbulent core leads to the formation of a single star 
surrounded by a disc, and the material in the disc then slowly accretes 
onto the star. A sink particle, representing the star and the inner
 part of the disc, is created where the density first exceeds 
$\rho_{_{\rm CRIT}} = 10^{-11}\,{\rm g\,cm}^{-3}$, and incorporating all 
the matter within $R_{_{\rm INIT}} = 5\,{\rm AU}$ of this point (Bate et 
al. 1995). (The values of $\rho_{_{\rm CRIT}}$ and $R_{_{\rm INIT}}$ are 
dictated by computational constraints. If $\rho_{_{\rm CRIT}}$ is 
increased, $R_{_{\rm INIT}}$ must be reduced, and more computational 
time is needed.)

\begin{figure}
\centerline{
\includegraphics[width=5cm]{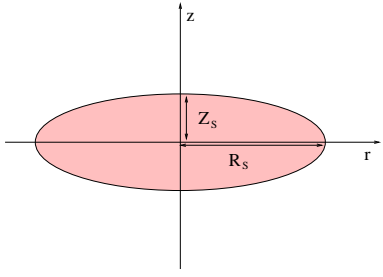}}
\caption{We use an oblate spheroidal sink particle, called a smartie, 
which contains a central star and a disc, and launches a bipolar outflow.}
\label{fig.smartie}
\end{figure}

We use a newly developed type of sink called a {\it smartie}, which is a 
rotating oblate spheroid with radius $R_{_{\rm S}}$ and half-thickness 
$Z_{_{\rm S}}$ (Fig.~\ref{fig.smartie}). The star is a point mass 
$M_{_\star}$ at the centre of the smartie, and the spheroid represents 
the inner part of the accretion disc surrounding the star. $R_{_{\rm S}}$ 
and $Z_{_{\rm S}}$ are calculated by assuming that the inner part of the 
accretion disc (the part inside the spheroid) has uniform density, 
temperature and angular speed, and that its internal energy and spin 
angular momentum balance its self-gravity and the gravity of the central 
star. 

The SPH particles which enter the smartie during a timestep, $(t,t+\Delta t)$, 
are assimilated by it at the end of the timestep, $t+\Delta t$, and their 
mass is initially added to the mass of the smartie, $M_{_{\rm S}}$, i.e. to 
the mass of the inner disc, 
\begin{equation} \label{EQN:ADDMASS}
M_{_{\rm S}}' \;=\; {M}_{_{\rm S}} \,+\, \sum_{i}\left\{m_i\right\} \,.
\end{equation}
At the same time, the centre of mass, ${\bf R}_{_{\rm S}}$, and velocity, 
${\bf V}_{_{\rm S}}$, of the smartie are adjusted to conserve linear momentum, 
\begin{eqnarray} \label{EQN:ADDPOS}
{\bf R}_{_{\rm S}}' & = & 
\frac{M_{_{\rm S}}{\bf R}_{_{\rm S}} \,+\, 
\sum_{i}\left\{ m_i {\bf r}_i \right\}}
{M_{_{\rm S}}'} \,, \\ \label{EQN:ADDVEL}
{\bf V}_{_{\rm S}}' & = & 
\frac{M_{_{\rm S}}{\bf V}_{_{\rm S}} \,+\, 
\sum_{i}\left\{ m_i {\bf v}_i \right\}}
{M_{_{\rm S}}'} \,,
\end{eqnarray}
and the intrinsic angular momentum (spin) of the smartie is adjusted 
to conserve angular momentum,
\begin{eqnarray} \nonumber
{\bf H}_{_{\rm S}}' & = &  {\bf H}_{_{\rm S}} \,+\, M_{_{\rm S}} 
\left({\bf R}_{_{\rm S}} - {\bf R}_{_{\rm S}}'\right) \otimes 
\left({\bf V}_{_{\rm S}} - {\bf V}_{_{\rm S}}'\right) \\ \label{EQN:ADDSPIN}
 & & \hspace{0.2cm} \,+\, \sum_{i} \left\{
m_i \left({\bf r}_i - {\bf R}_{_{\rm S}}'\right) \otimes 
\left({\bf v}_i -{\bf V}_{_{\rm S}}'\right) \right\} \,.
\end{eqnarray}
In Eqns, \ref{EQN:ADDMASS} through \ref{EQN:ADDSPIN}, primed variables 
($M_{_{\rm S}}'$, ${\bf R}_{_{\rm S}}'$, ${\bf V}_{_{\rm S}}'$, 
${\bf H}_{_{\rm S}}'$) are used to denote sink parameters adjusted for 
the addition of assimilated SPH particles.

The smartie mass evolves according to
\begin{equation} \label{EQN:MASS}
\frac{dM_{_{\rm S}}}{dt} \;=\; \left. \frac{dM}{dt} \right|_{_{\rm IN}} 
\,-\, \frac{dM_{_\star}}{dt} \,-\, \left. \frac{dM}{dt} \right|_{_{\rm OUT}} \,,
\end{equation}
where the first term on the right of Eqn.~\ref{EQN:MASS} represents SPH particles 
assimilated by the smartie, as discussed in the preceding paragraph. 

The second term on the right of Eqn.~\ref{EQN:MASS} 
represents accretion onto the central star, and is given by
\begin{eqnarray}
\frac{dM_{_\star}}{dt} & = & -\,\frac{0.9 M_{_{\rm S}}}{t_{_{\rm VISC}}} \,, \\
t_{_{\rm VISC}} & = & 
\frac{\left[G(M_{_\star} + M_{_{\rm S}})R_{_{\rm S}}\right]^{1/2}}
{\alpha_{_{\rm SS}}\,a_{_{\rm S}}^2} \, \left\{ 1 \,+\, \frac{G M_{_{\rm S}}^2}
{(M_{_\star} + M_{_{\rm S}}) R_{_{\rm S}} a_{_{\rm S}}^2} \right\}^{-1} \,,
\end{eqnarray}
where 
$\alpha_{_{\rm SS}}$ is the standard Shakura \& Sunyaev (1973) turbulent 
viscosity parameter and $a_{_{\rm S}}$ is the mean sound speed in the smartie. 
The leading term represents turbulent viscosity, and the term in braces 
adjusts this for angular momentum transport by gravitational torques in 
a Toomre-unstable disc (cf. Lin \& Pringle 1990).

The third term on the right of Eqn.~\ref{EQN:MASS} represents mass loss from 
the smartie, which can occur in two ways. (i) SPH particles are created at 
rate $0.1 M_{_{\rm S}} / t_{_{\rm VISC}}$ and launched along the rotation 
axis with speed $100\,{\rm km\,s}^{-1}$ to form a bipolar outflow. These jets 
push the surrounding material aside and create hourglass cavities about the 
rotation axis. (ii) In the present model the outflow does not remove any angular 
momentum. Therefore the redistribution of angular momentum which drives accretion 
onto the central star causes the residual inner disc (which has to carry all 
the angular momentum assimilated by the smartie) to expand. When this happens, 
the smartie is allowed to expand until $R_{_{\rm S}} = 16\,{\rm AU}$. 
Thereafter, $R_{_{\rm S}}$ is held constant by excreting SPH particles -- and 
with them angular momentum -- just outside the equator of the smartie.

The luminosity of the star is given by
\begin{equation} \label{EQN:LUM}
L_{_\star} \;=\; {\rm L}_{_\odot} \left( \frac{M_{_\star}}{{\rm M}_{_\odot}} 
\right)^3 \,+\, \frac{G M_{_\star}}{R_{_\star}} \frac{dM_{_\star}}{dt} \,,
\end{equation}
where $M_{_\star}$ and $R_{_\star}$ are the mass and radius of the star, and 
we set $R_{_\star} = 3 {\rm R}_{_\odot}$ (Stahler 1988). 

The mean smartie temperature, and hence also the mean smartie sound speed, are 
determined by the balance between radiative cooling, heating by radiation from 
the star and from the background radiation field, and viscous dissipation in 
the inspiralling gas of the smartie. A full description of the implementation 
of smarties is given in Boyd (2003), and the consequences of feedback from 
bipolar outflows are explored in detail in Boyd et al. (in preparation).

\section{The radiative transfer simulation}

For the radiative transfer simulations, we use a version of {\sc phaethon}, 
a Monte Carlo radiative transfer code which we have developed (Stamatellos 
\& Whitworth 2003) and optimised for radiative transfer simulations on SPH 
density fields (Paper I). The main developments of the optimised version 
are (i) that it uses the tree structure inherent within the SPH code to 
construct a grid of cubic radiative transfer (RT) cells, spanning the 
entire computational domain, the {\it global grid}; and (ii) that in 
addition it constructs a local grid of concentric spherical RT cells 
around each star, a {\it star grid}, to capture the steep temperature 
gradients that are expected in the vicinity of a star. The code 
reemits luminosity packets are soon as they are absorbed (Lucy 1999) and
uses the frequency distribution adjustment technique developed by 
Bjorkman \& Wood (2000). 
 
\subsection{Construction of the global radiative transfer grid}

To construct RT cells for a given time-frame from an SPH simulation, we 
take advantage of the fact that SPH uses an octal tree structure (to group 
particles together when calculating gravity forces and also to find 
neighbours; for details see Paper I). The SPH tree is a recursive 
hierarchical division of the computational domain into cubic cells within 
cubic cells (Barnes \& Hut 1986). When the SPH tree is being built, we 
record information about the size of each cell and the number of SPH 
particles it contains. The RT cells of the global grid are then the largest 
SPH cells which contain $\leq N_{_{\rm MAX}}$ particles. This means that 
the RT cells automatically have comparable mass. We choose $N_{_{\rm MAX}} 
\sim 100\;{\rm to}\;150$, i.e. somewhat larger than the mean number of SPH 
neighbours, $N_{_{\rm NEIB}} \sim 50$. Consequently the size of the RT 
cells is on the order of the smoothing length, $h$, and so the temperature 
resolution is similar to the density resolution of the SPH simulation.

In order to increase the resolution of the radiative transfer simulation, 
we use particle splitting (Kitsionas \& Whitworth 2002). This method 
replaces each SPH particle (parent particle) with a small group of 13 
particles (children particles), each one having 1/13 of the mass of the 
parent particle. The children are placed on an hexagonal close-packed 
array, with one of them in the centre of the array and the other 12 
equidistant from the first one. By doing this the smoothing length is 
decreased by a factor of $13^{-1/3}=0.425$,  and the average size of the 
radiative transfer cells is reduced by the same factor.

\subsection{Construction of the local radiative transfer grid around a star}

\begin{figure}
\centerline{
\includegraphics[width=8.5cm]{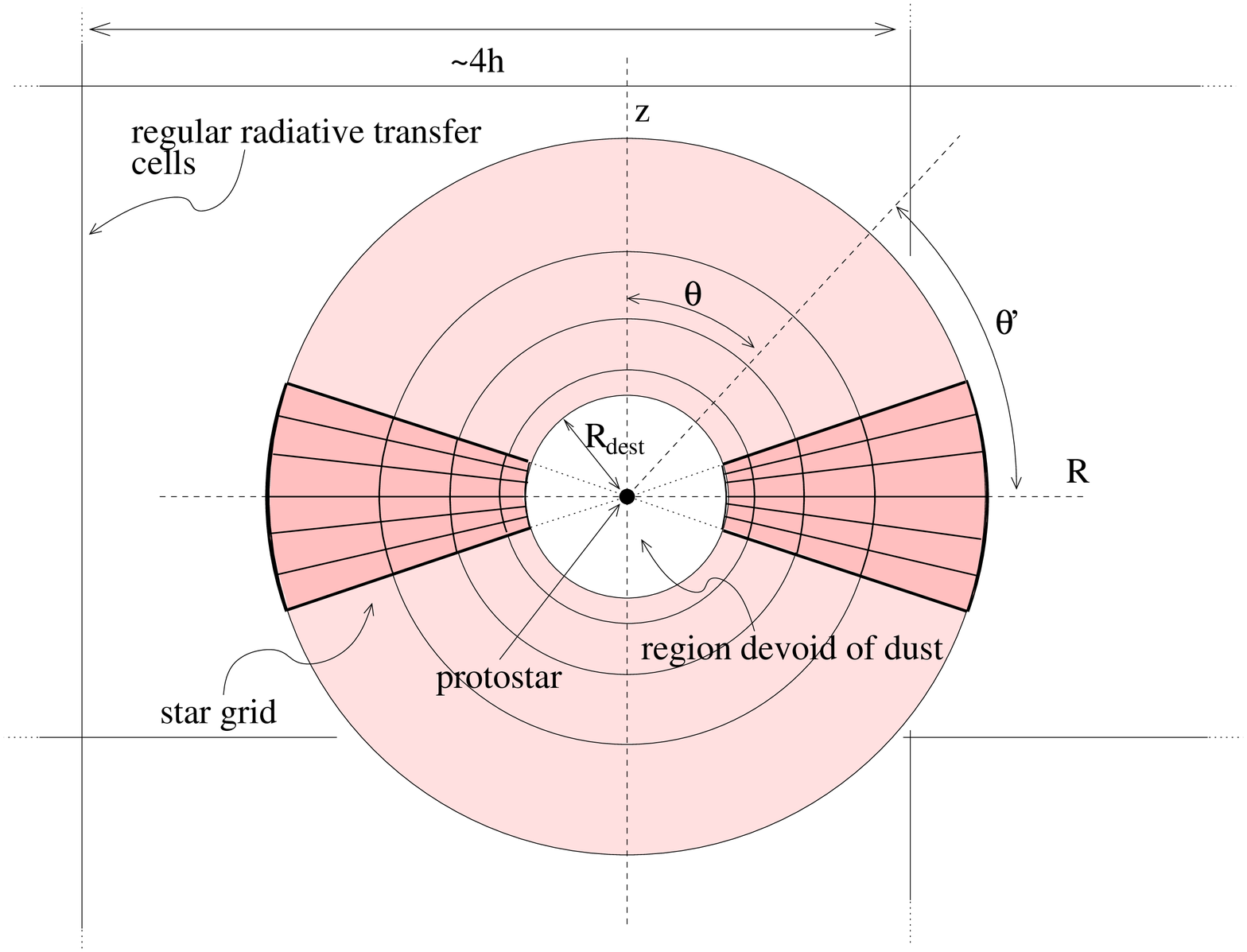}}
\caption{The star grid consists of concentric spherical surfaces centred 
on the star, with equal spacing in $\log (r)$, and conical surfaces 
symmetric about the angular momentum of the smartie, with equally spaced 
opening angle $\theta$. This grid takes account of the dust-free cavity 
in the immediate vicinity of the star, and resolves the disc inside the 
smartie.}
\label{fig.smartie.rt}
\end{figure}

In Paper I we discussed the need for a high-resolution grid of radiative 
transfer cells  -- a {\it star grid} -- in order to capture the steep 
temperature gradients that are expected in the vicinity of a star. The star 
grid discussed there was developed for treating the unresolved region inside 
a spherically symmetric sink, and so a one-dimensional star grid consisting 
of concentric spherical cells centred on the star was sufficient. 

In the case we examine here the sinks are smarties, i.e. oblate spheroids, 
and the unresolved region inside a smartie represents both the protostar and 
the inner protostellar disc. Thus a spherically symmetric one-dimensional star grid 
is inappropriate. Instead we adopt a similar approach to Kurosawa et al. 
(2004) and represent the unresolved region near the star with a flared disc 
(Fig.~\ref{fig.smartie.rt}).

However, the properties of the inner disc are not imposed externally, as in 
Kurosawa et al. (2004), but are determined by the properties of the smartie, 
and therefore attempt to capture the physics of a protostellar disc accreting 
onto a newly-formed protostar (as outlined in Section~\ref{S:SPH}). 
Specifically, we adopt an inner disc with  inner radius $R_{_{\rm DEST}}$, outer 
radius $R_{_{\rm SG}}$, and constant ratio of thickness, $H$, to radius, $r$,
\begin{equation}
H(r) \;=\; \alpha\,r\,,\;\;\;\;{\rm where}\;\;\;\;
\alpha \;=\; \frac{Z_{_{\rm S}}}{R_{_{\rm S}}} \,.
\end{equation}

The dust destruction radius is given by
\begin{equation}
R_{_{\rm DEST}} \;=\; \frac{R_{_\star}}{2}\,\left( \frac{T_{_\star}}
{T_{_{\rm DEST}}} \right)^{(4+\beta)/2}\,,
\end{equation}
where $\beta$ is the dust emissivity index (see Paper I). There is no dust 
inside $R_{_{\rm DEST}}$.

We put  $R_{\rm SG} = 16\,{\rm AU}$, since this is the maximum size of a 
smartie, and it is comparable with the size of the local global RT cells 
(the cubic cells constructed from the SPH tree).

For $z < 4 \alpha r$ and $R_{_{\rm DEST}} < r < R_{_{\rm SG}}$, the density 
of the inner disc is given by 
\begin{equation}
\rho (z,r)= \rho_{_{\rm DEST}}\,\left(\frac{r}{R_{_{\rm DEST}}}\right)^{-2}\,
\exp\left[ -\,\frac{1}{2}\left( \frac{z}{\alpha r} \right)^2 \right]\,,
\end{equation}
(cf. Wood et al. 2002a, 2002b). $\rho_{_{\rm DEST}}$ is fixed by equating 
the mass of this inner disc to the mass of the disc inside the smartie. For 
$z > 4 \alpha r$ and $R_{_{\rm DEST}} < r < R_{_{\rm SG}}$, the density 
is set equal to the average density of the local global RT cells. 

The inner disc is then divided into RT cells by spherical and conical surfaces 
(see Fig.~\ref{fig.smartie.rt}). The spherical surfaces are evenly spaced in 
logarithmic radius and there are typically 20 of them. The conical surfaces 
are evenly spaced in polar angle and there are typically 30 of them.

\subsection{Radiation sources} 
\label{SS:RADSOURCES}

\begin{figure}
\centerline{
\includegraphics[width=7.5cm]{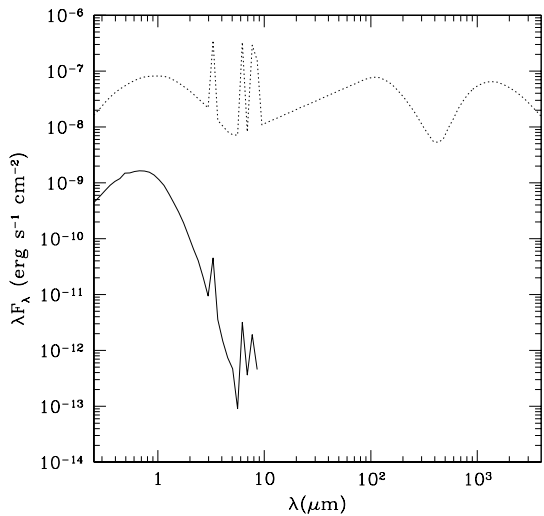}}
\caption{Dotted line: The external illumination field 
(BISRF + emission from PAHs).
Solid line: The radiation scattered by the molecular cloud. 
This is similar for all
the time-frames and viewing angles, for the models we examine here.}
\label{fig.scat}
\end{figure}
\begin{table*}
\begin{center}
\small\
\caption{Model parameters}
\begin{tabular}{@{}|c|c|c|c|c|c|c|c|c|l|} \hline
id & time (yr) & $M_\star ({\rm M}_{\sun})$ 
& $R_{\rm S}$ (AU) & $Z_{\rm S}$ (AU)& $M_{\rm S} ({\rm M}_{\sun})$ 
& $\dot{M}_\star ({\rm M}_{\sun}$/yr) 
& $L_{\rm TOT} ({\rm L}_{\sun})$ & $T_\star$ (K) 
& description \\
\hline
{\texttt t0} & 0 & - & - &  - & - & - & - & - &initial conditions\\
\texttt{t1} & $5.3\times10^{4}$ & - & - &  - & - & - & - & - & before sink formation \\
\texttt{t2} & $5.3\times10^{4}$ & 0.01 & 4.0 &  3.4 & 0.04 & $5\times10^{-5}$ & 5.7  & 5150   & a protostar has formed \\
\texttt{t3} & $6.0\times10^{4}$ & 0.20  & 4.0 &  0.4 & 0.09 & $1\times10^{-5}$& 27.2 & 7650 & \\
\texttt{t4} & $6.8\times10^{4}$ & 0.49 & 4.0 &  0.4 & 0.05 & $2\times10^{-6}$& 12.3 & 6250 &\\
\texttt{t5} & $6.9\times10^{4}$ & 0.53 & 4.0 &  0.4 & 0.01 & $4\times10^{-7}$& 2.5  & 4200 & protostar is off centre \\
\hline
\end{tabular}\label{tab:model.params}
\end{center}
\hspace{0.8cm}
\begin{tabular}{@{}lll}
$Z_{\rm S}$, $R_{\rm S}$: Smartie dimensions (Fig.~\ref{fig.smartie}) &  $M_{\rm \star}$: Stellar mass&
$T_\star$: Star temperature\\
$M_{\rm S}$: Smartie mass  & $\dot{M}_\star$: Accretion rate onto the central star
&$L_{\rm TOT}$: Total star luminosity (intrinsic +accretion) \\
\end{tabular}
\end{table*}

The core is illuminated externally by the interstellar radiation field and 
internally by the newly formed protostar (once it has formed).

For the external radiation field  we adopt a revised version of the Black 
(1994) interstellar radiation field (BISRF). The BISRF consists 
of radiation from giant stars and dwarfs, thermal emission from dust grains, 
cosmic background radiation, and mid-infrared emission from transiently heated 
small PAH grains (Andr\'e et al. 2003), as illustrated on Fig.~\ref{fig.scat}. 
Typically the BISRF is represented by $10^9-10^{10}$ 
$L$-packets (see Paper I).

The protostar luminosity, $L_{_\star}$, and effective temperature, $T_{_\star}$ 
are given by Eqn.~\ref{EQN:LUM}; a blackbody spectrum at $T_{_\star}$ is assumed. 
For the systems examined here, the accretion contribution to the luminosity 
dominates, because the mass of the protostar is small ($<0.6\,{\rm M}_{\sun}$) 
and the accretion rate is high ($>10^{-7}\,{\rm M}_{\sun}$/yr). Typically the 
protostar is represented by $10^6\;{\rm to}\;10^8$ $L$-packets.

\subsection{Dust opacity}

Like other studies of prestellar cores (e.g. Evans et al. 2001, Young et 
al. 2004) we use the Ossenkopf \& Henning (1994) opacity for a standard 
MRN interstellar grain mixture (53\% silicate and 47\% graphite) which has 
coagulated and accreted thin ice mantles over a period of $10^5$ yr at a 
density of $10^6\; {\rm cm}^{-3}$. However, we emphasize that the opacity 
of the dust in cores is very uncertain (e.g. Bianchi et al. 2003).

\section{The evolution of the dust temperature field, SED and isophotal maps
in a star-forming core}

We perform radiative transfer simulations on 6 time-frames during the collapse 
of a star-forming core. We focus our attention just before and just after the 
formation of the first protostar in the core. The characteristics of each 
time-frame are listed in Table~\ref{tab:model.params}. The simulations are 
3-dimensional and provide dust temperature fields, SEDs and isophotal maps.

\begin{figure*}
\centerline{
\includegraphics[width=9.9cm]{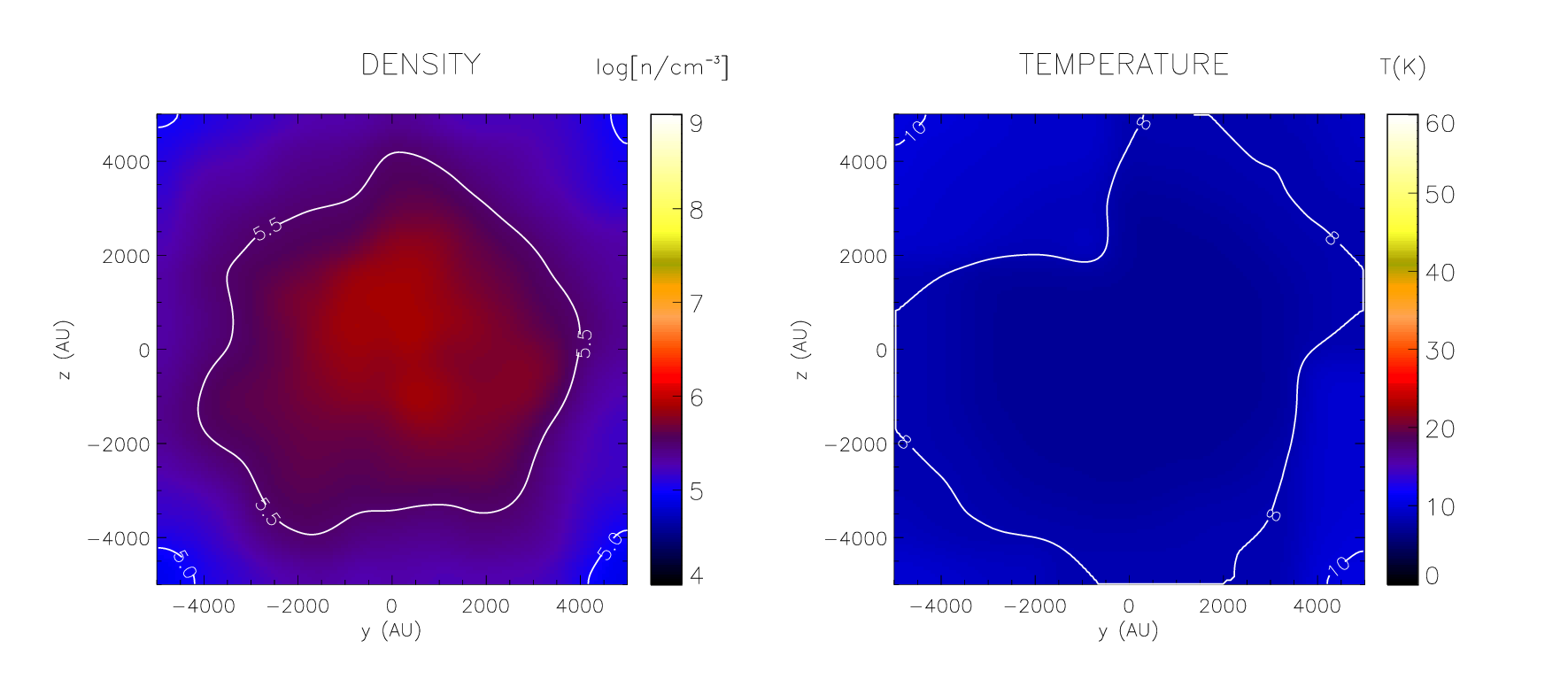}
\hspace{-1cm}
\includegraphics[width=9.9cm]{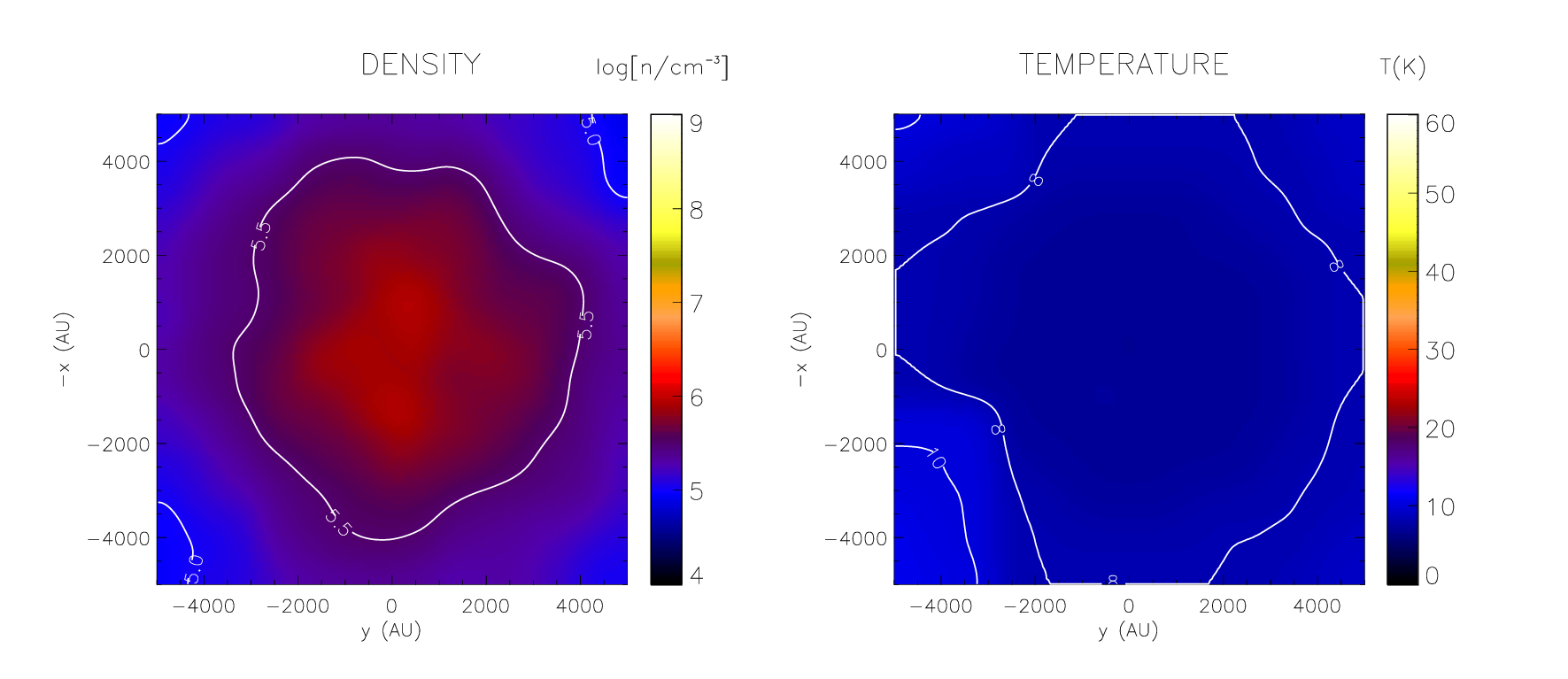}}\vspace{-0.4cm}
\centerline{
\includegraphics[width=9.9cm]{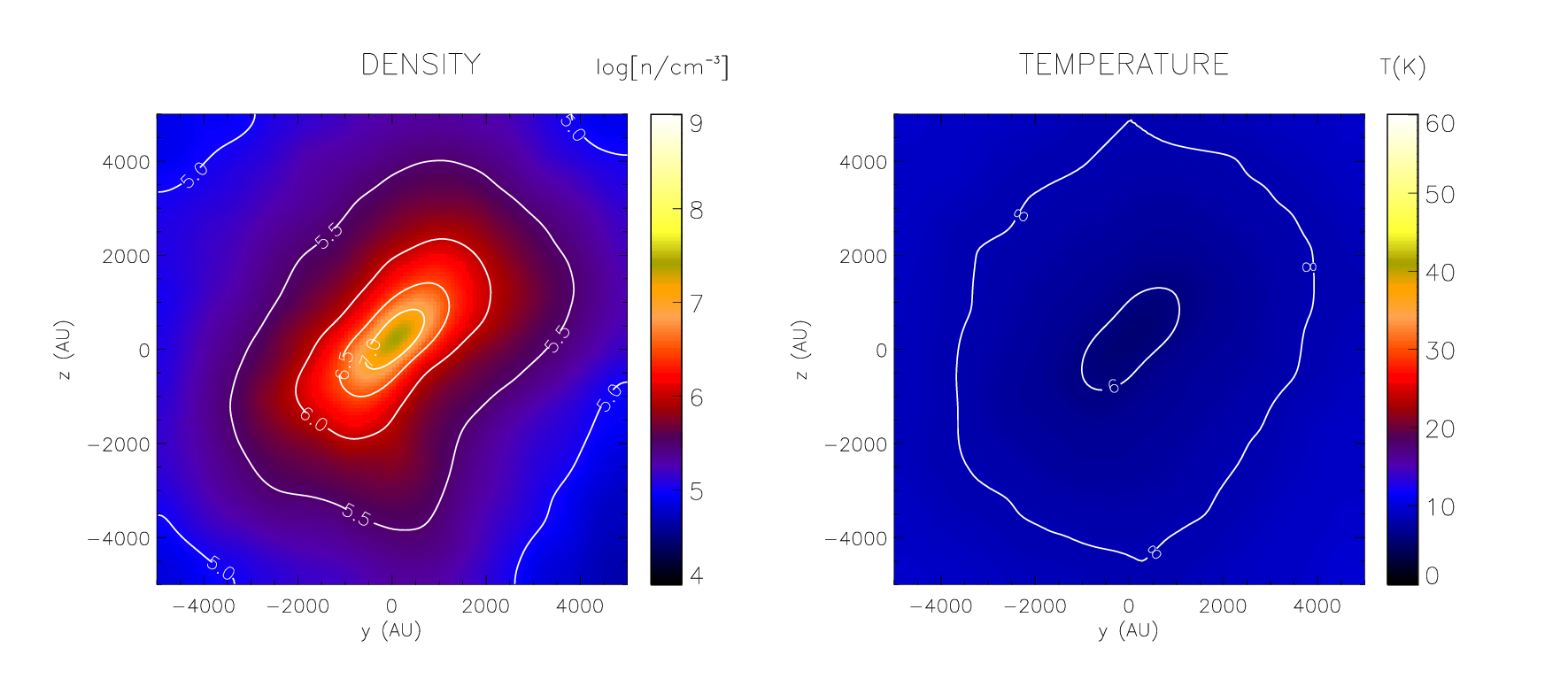}
\hspace{-1cm}
\includegraphics[width=9.9cm]{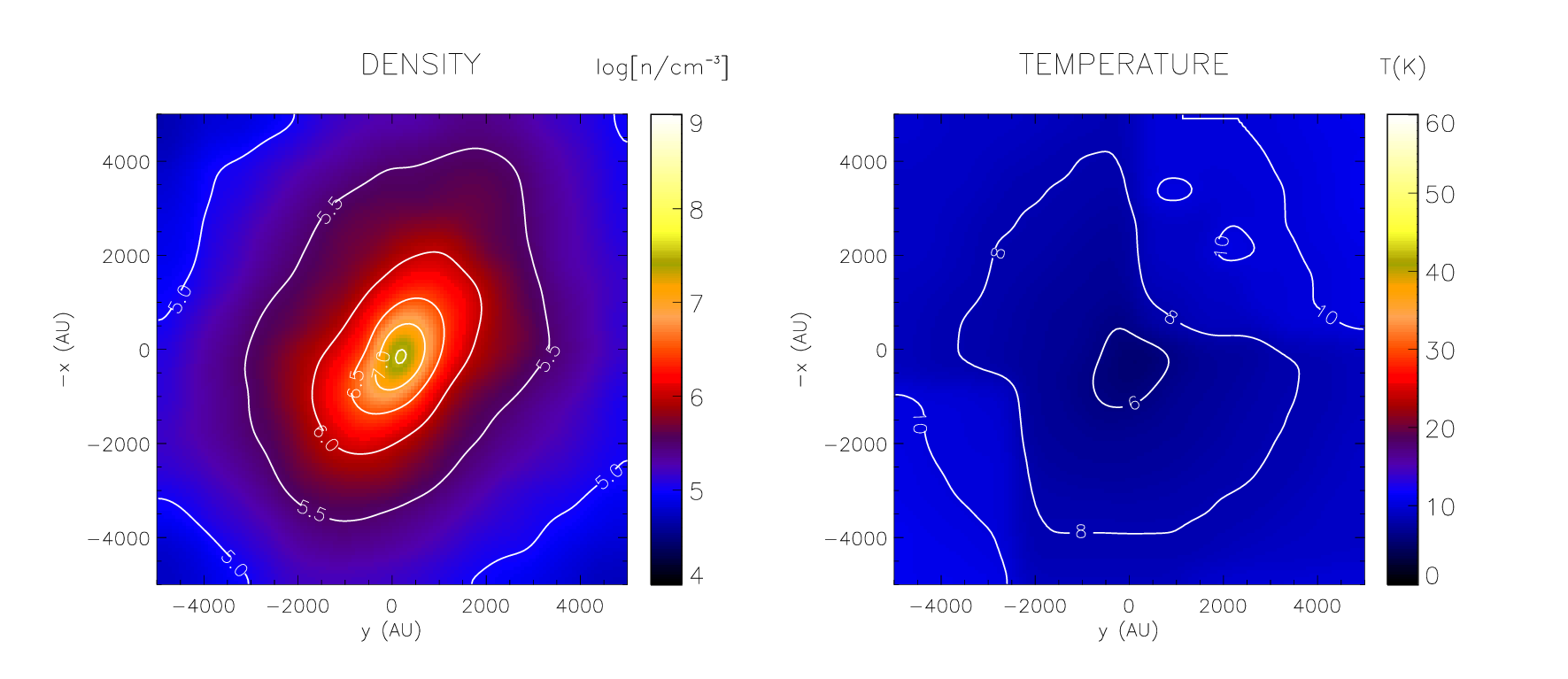}}\vspace{-0.4cm}
\centerline{
\includegraphics[width=9.9cm]{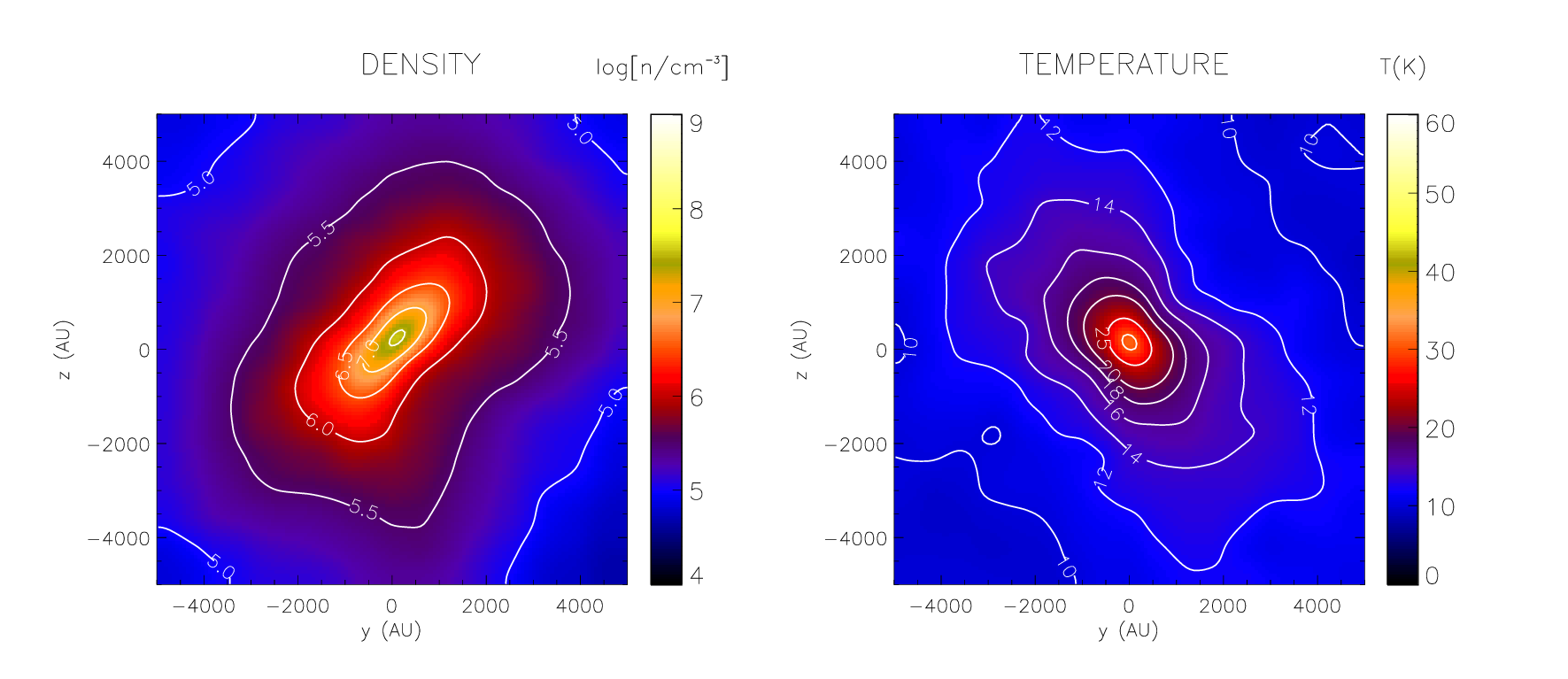}
\hspace{-1cm}
\includegraphics[width=9.9cm]{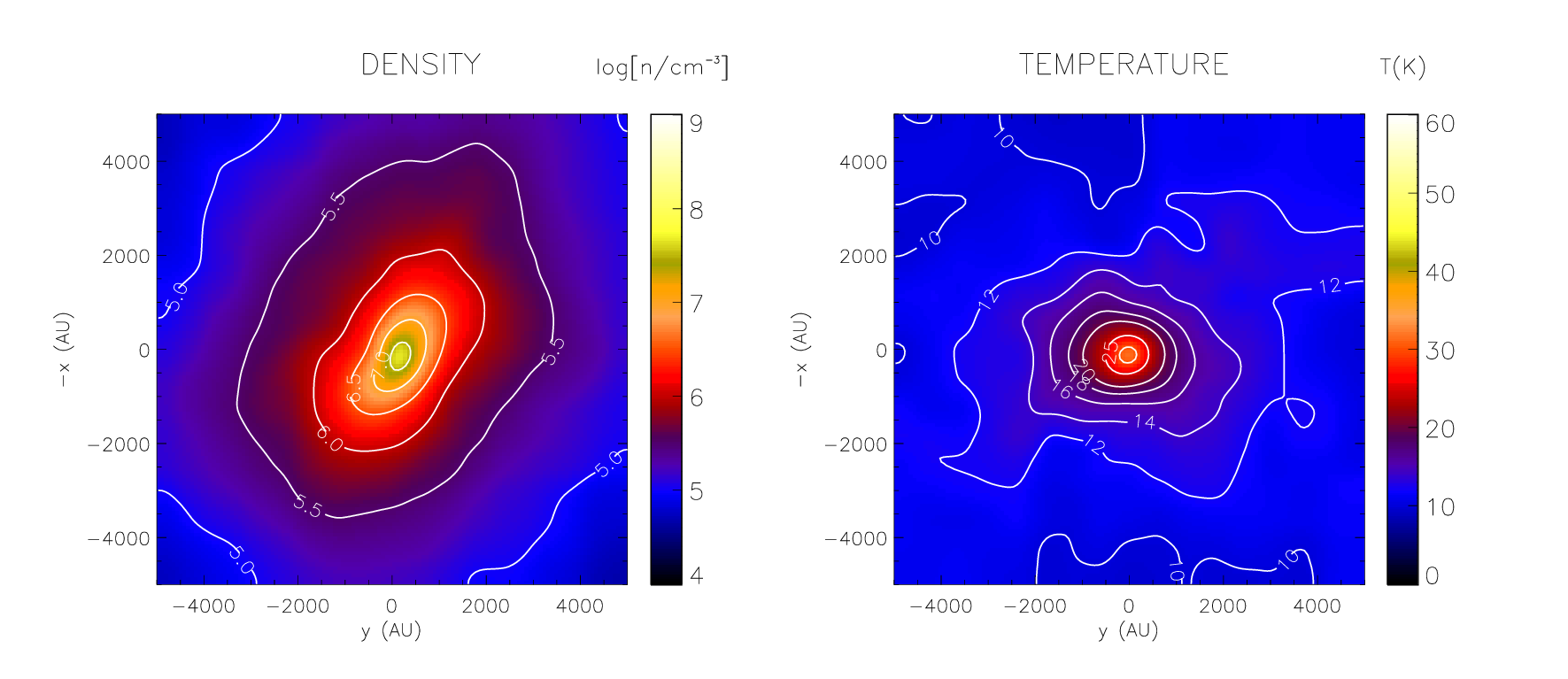}}\vspace{-0.4cm}
\centerline{
\includegraphics[width=9.9cm]{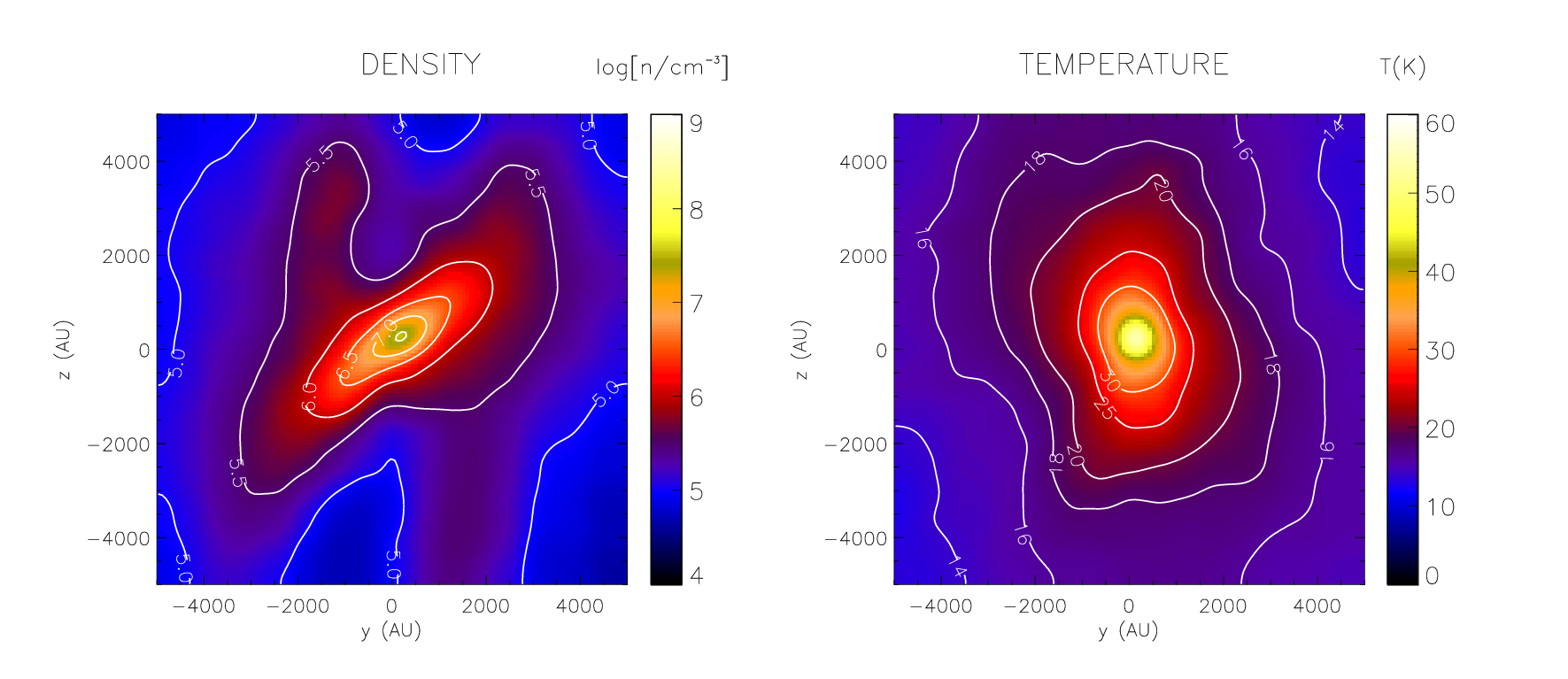}
\hspace{-1cm}
\includegraphics[width=9.9cm]{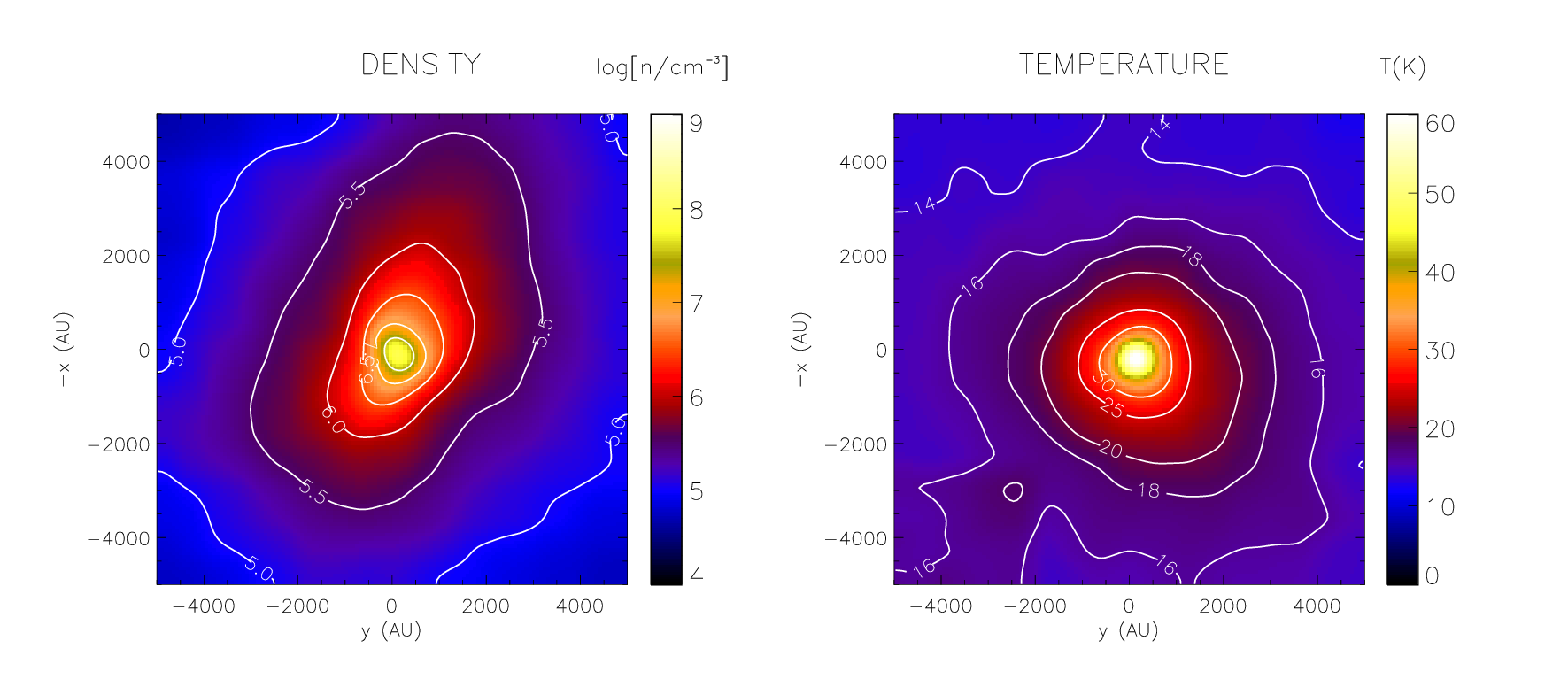}}\vspace{-0.4cm}
\centerline{
\includegraphics[width=9.9cm]{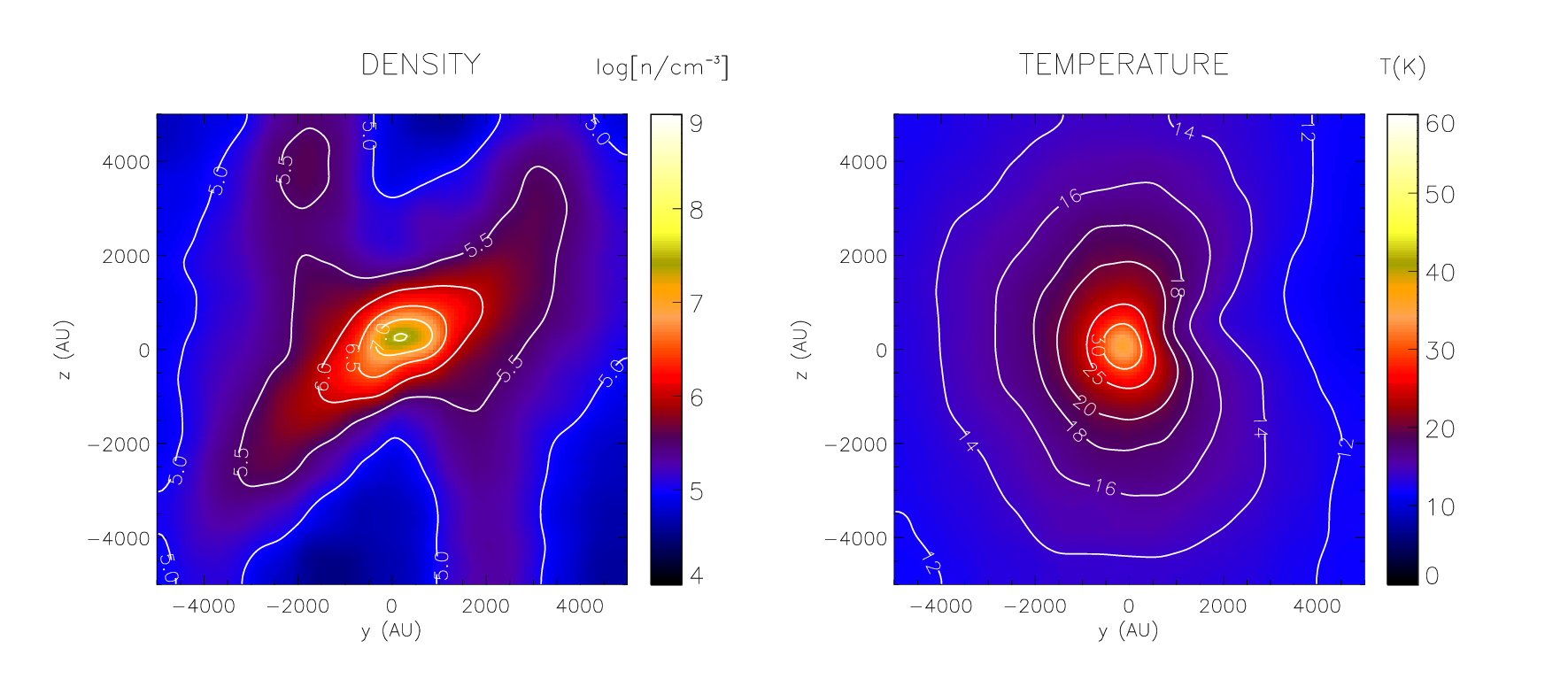}
\hspace{-1cm}
\includegraphics[width=9.9cm]{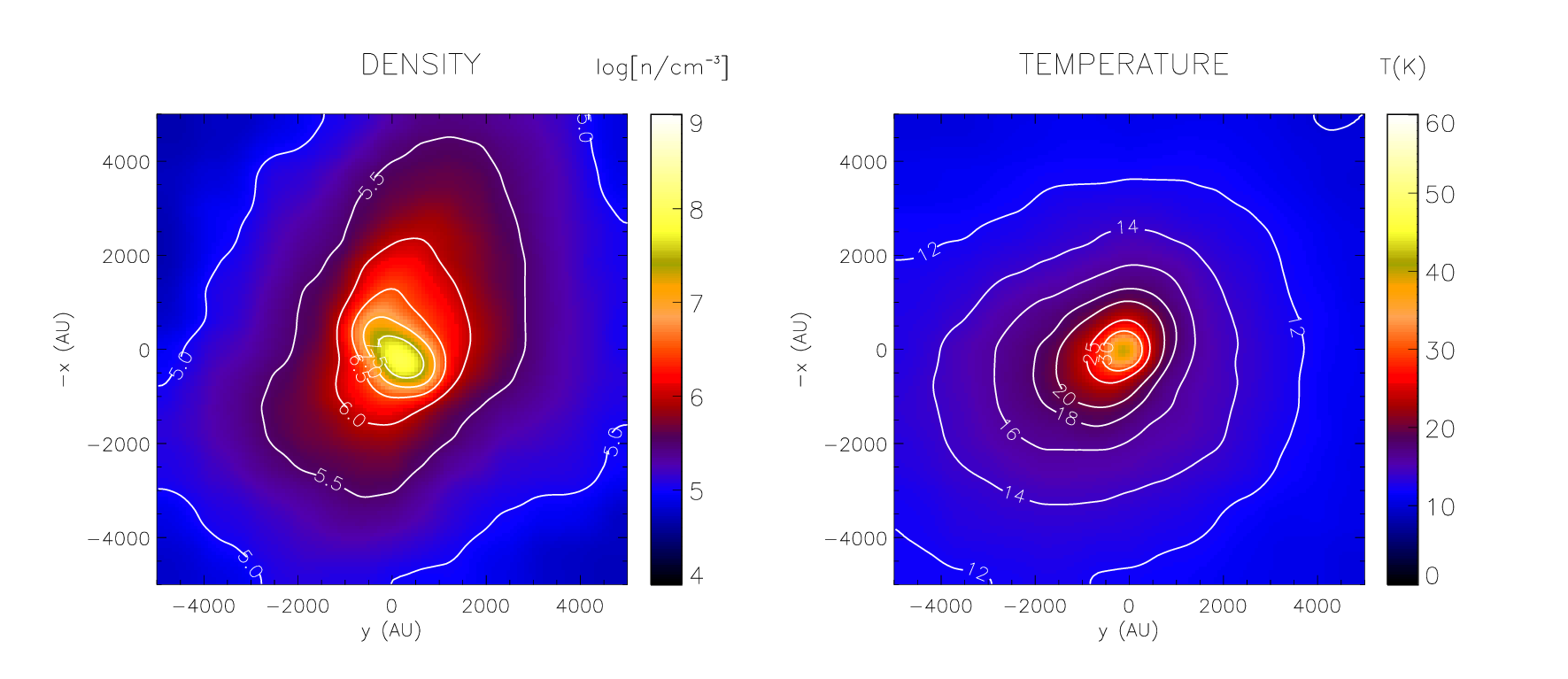}}\vspace{-0.4cm}
\centerline{
\includegraphics[width=9.9cm]{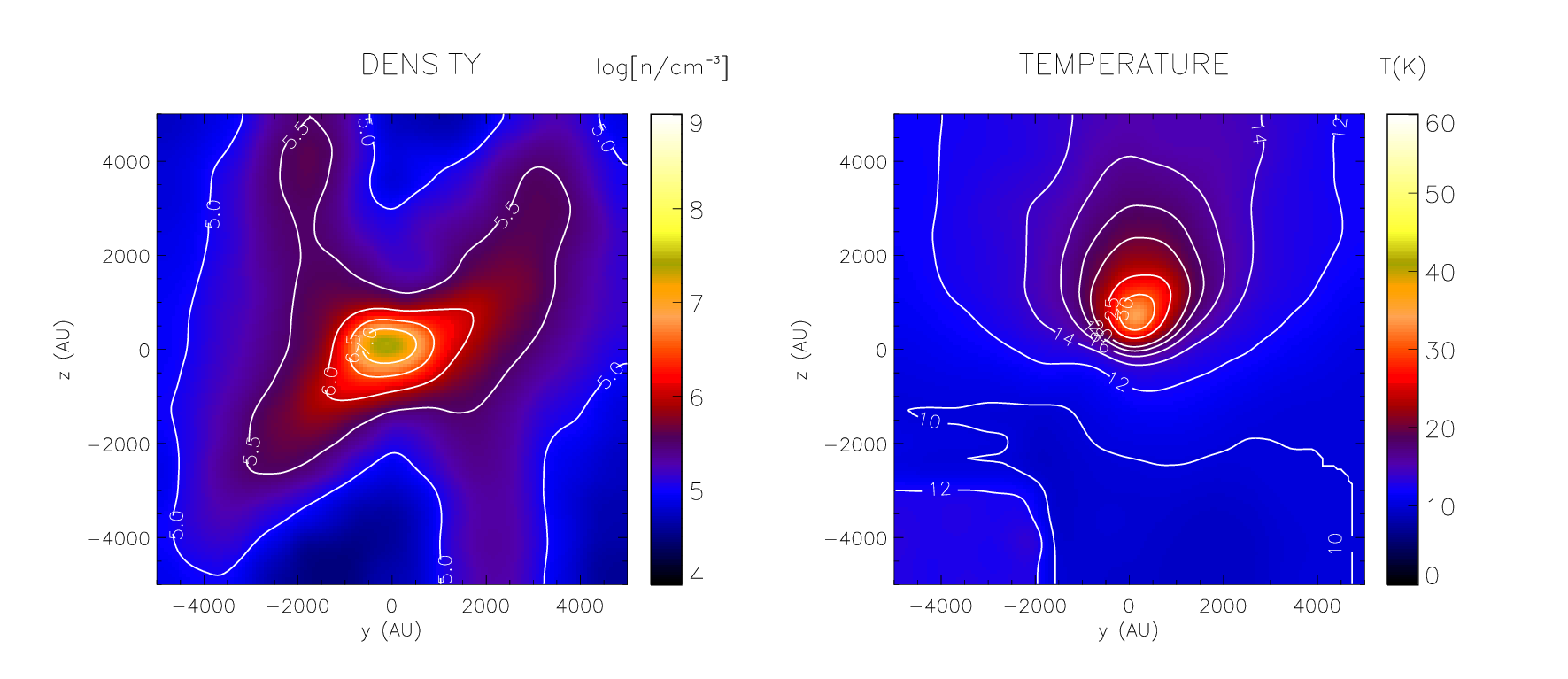}
\hspace{-1cm}
\includegraphics[width=9.9cm]{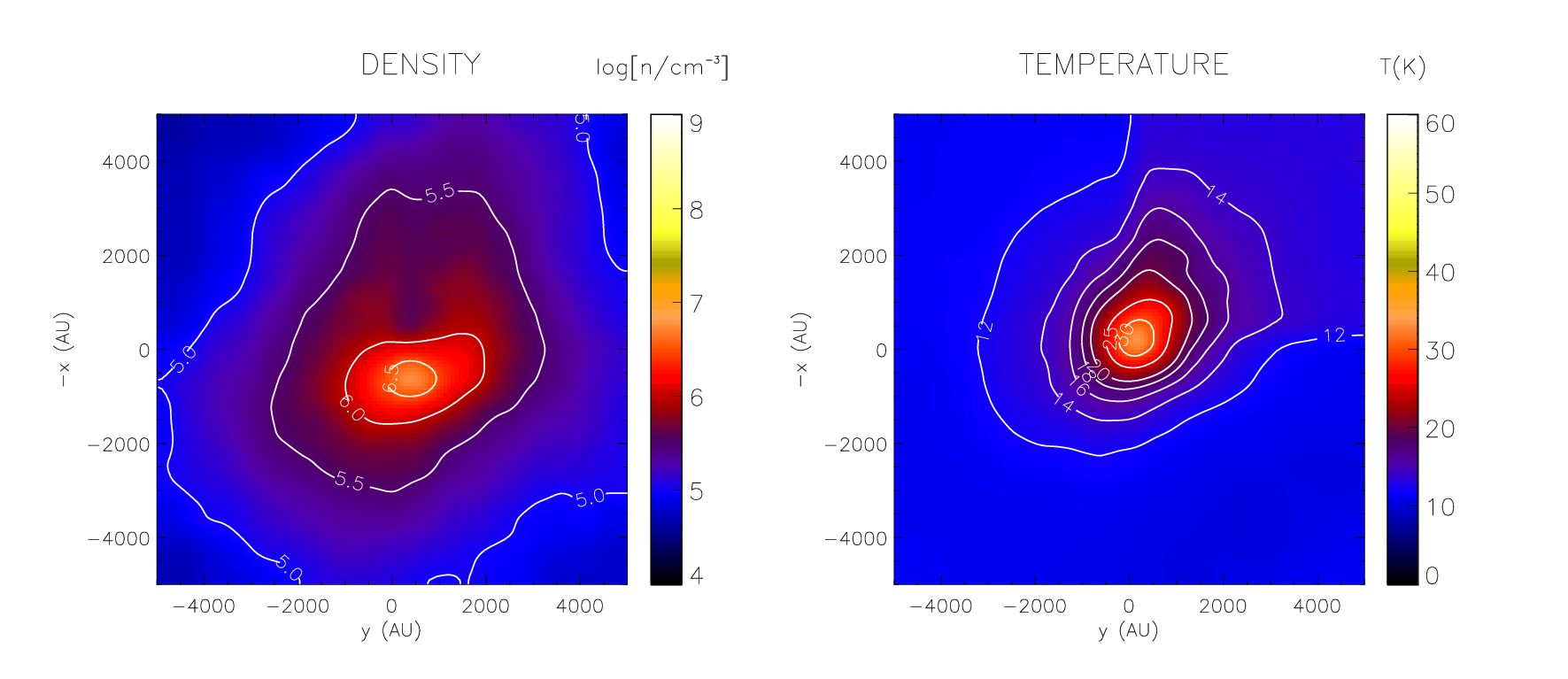}}\vspace{-0.3cm}
\caption{Cross sections of density and dust temperature on a plane 
parallel to the $x=0$ plane (left two columns) and parallel to the 
$z=0$ plane (right two columns). Each row corresponds to a different 
time frame from Table~\ref{tab:model.params}, top to bottom {\texttt 
t0} to {\texttt t5}. The planes are chosen so as to include the 
maximum density region ({\texttt t0} and {\texttt t1}) or the 
protostar ({\texttt t2} to {\texttt t5}).}
\label{fig.dens.temp}
\end{figure*}

\subsection{Dust temperature fields in prestellar cores and young protostars}

\begin{figure}
\centerline{
\includegraphics[width=4.2cm]{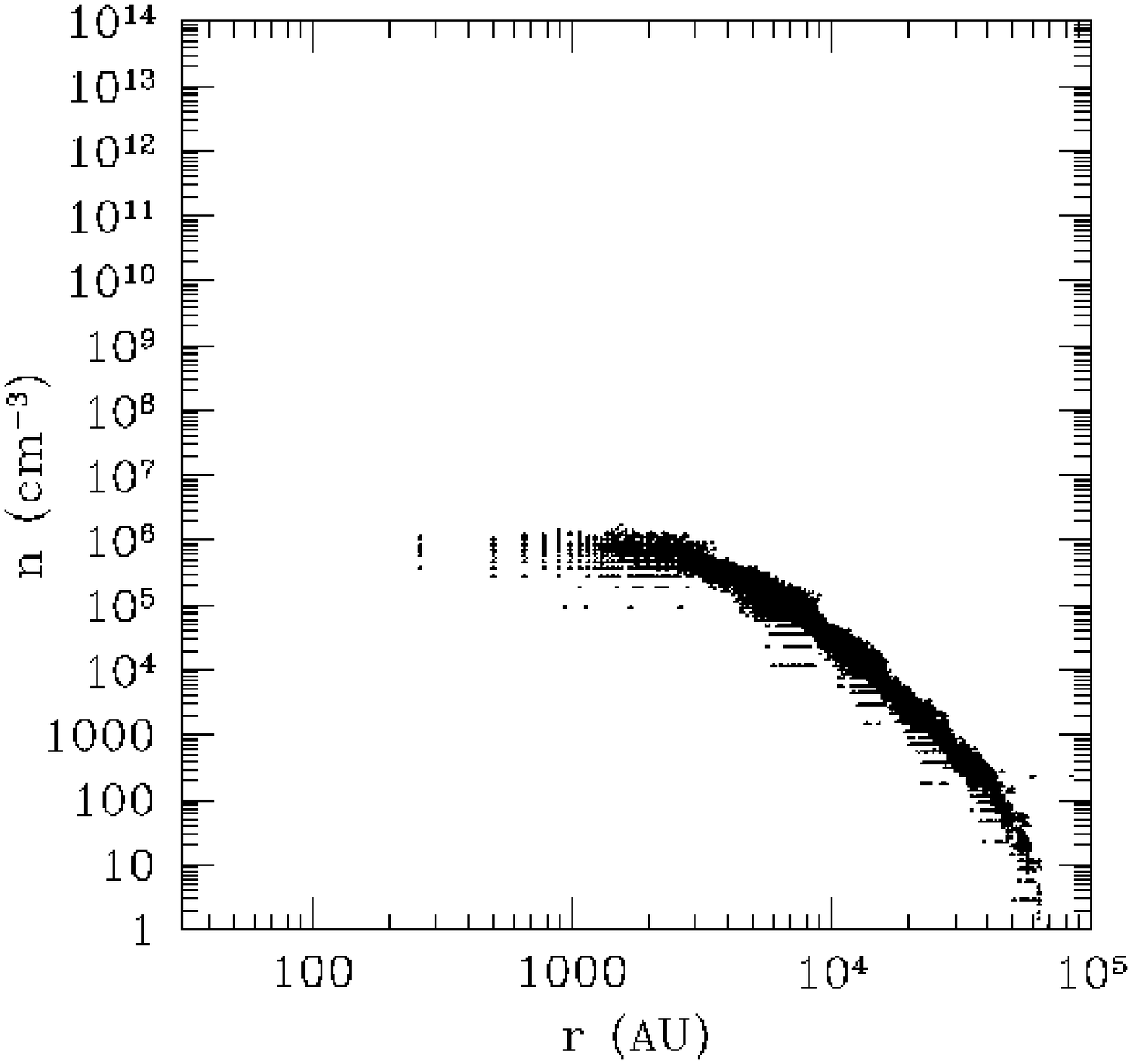}
\includegraphics[width=4.2cm]{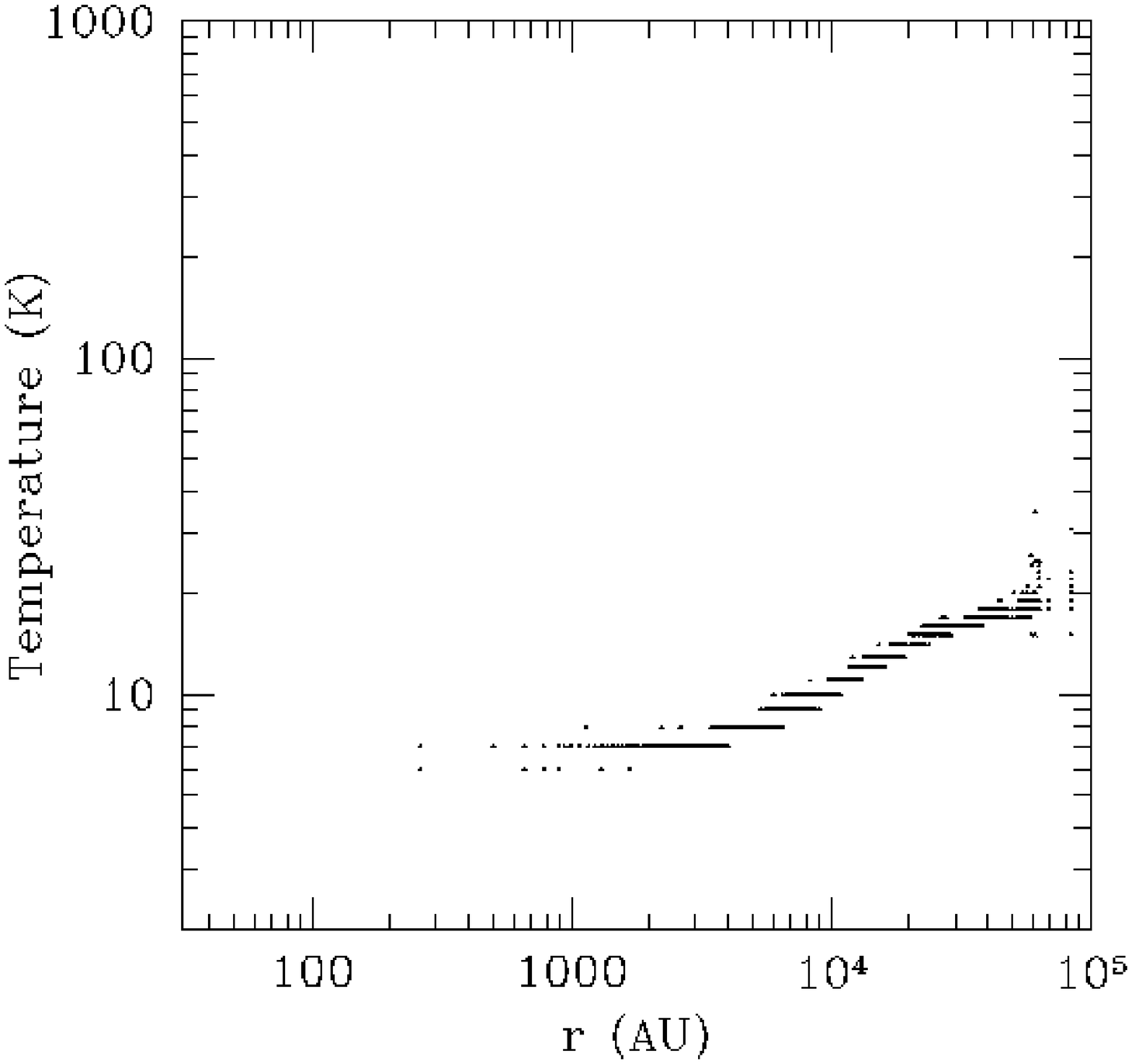}
}\vspace{-0.5cm}
\centerline{
\includegraphics[width=4.2cm]{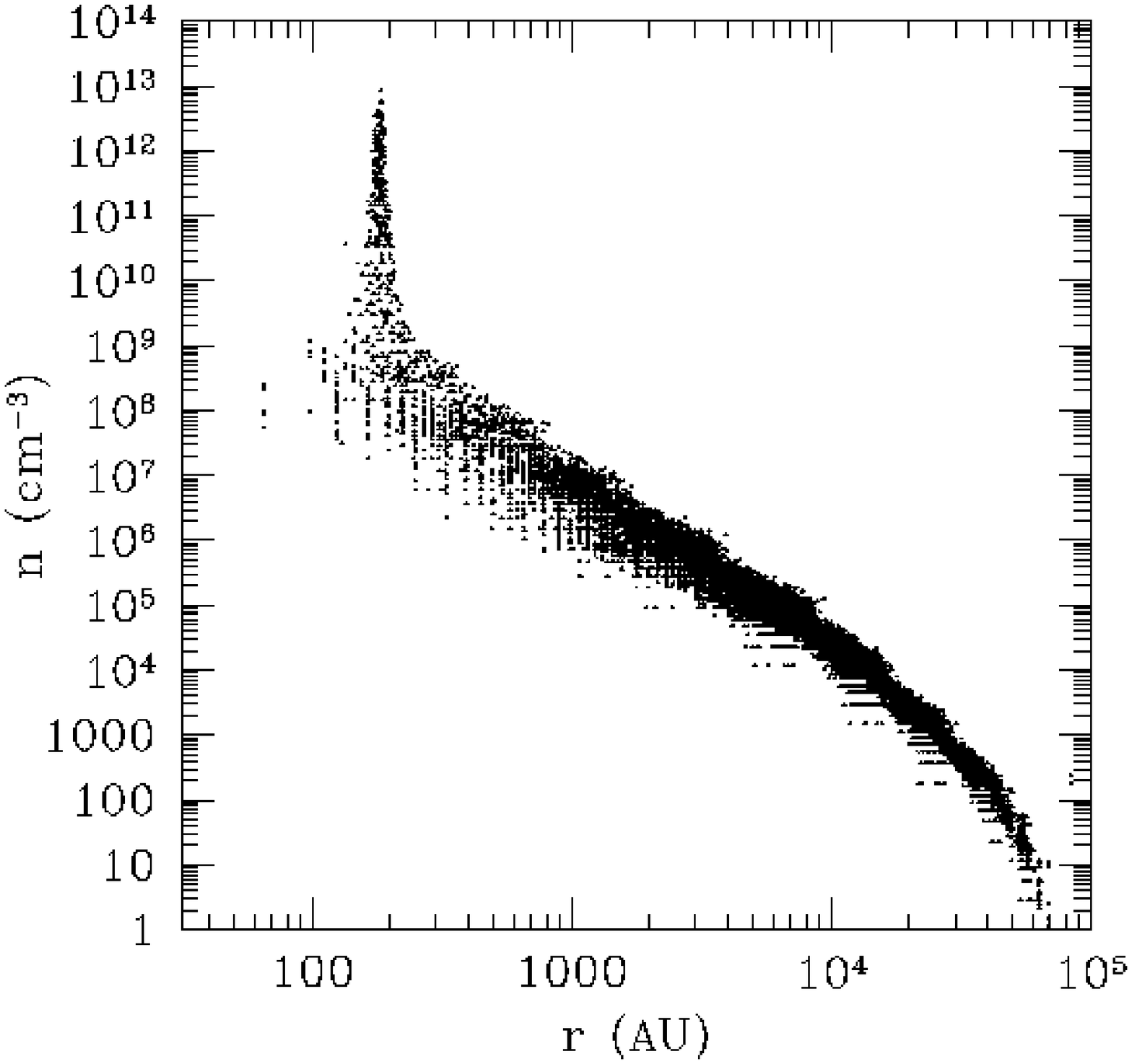}
\includegraphics[width=4.2cm]{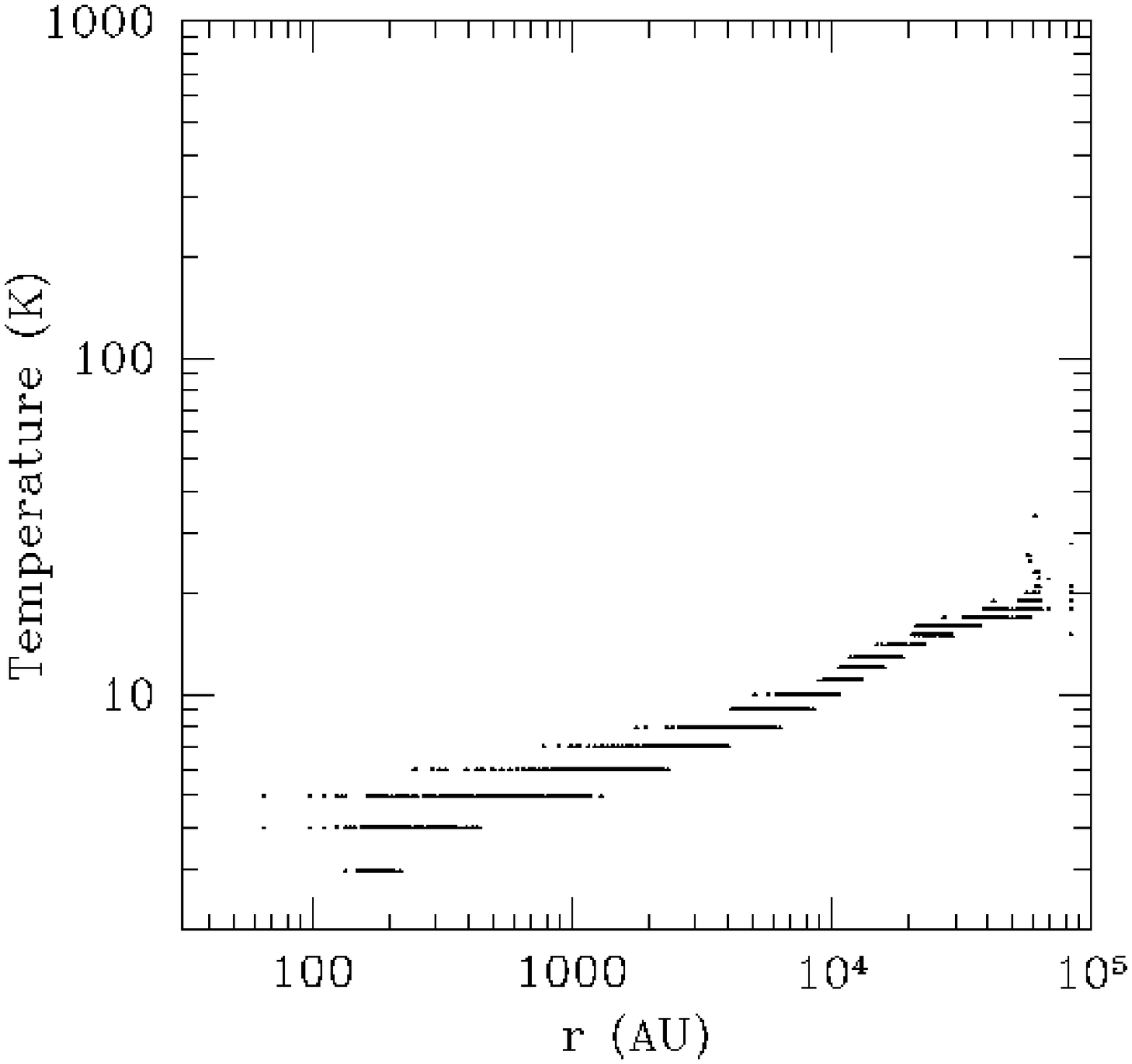}
}\vspace{-0.5cm}
\centerline{
\includegraphics[width=4.2cm]{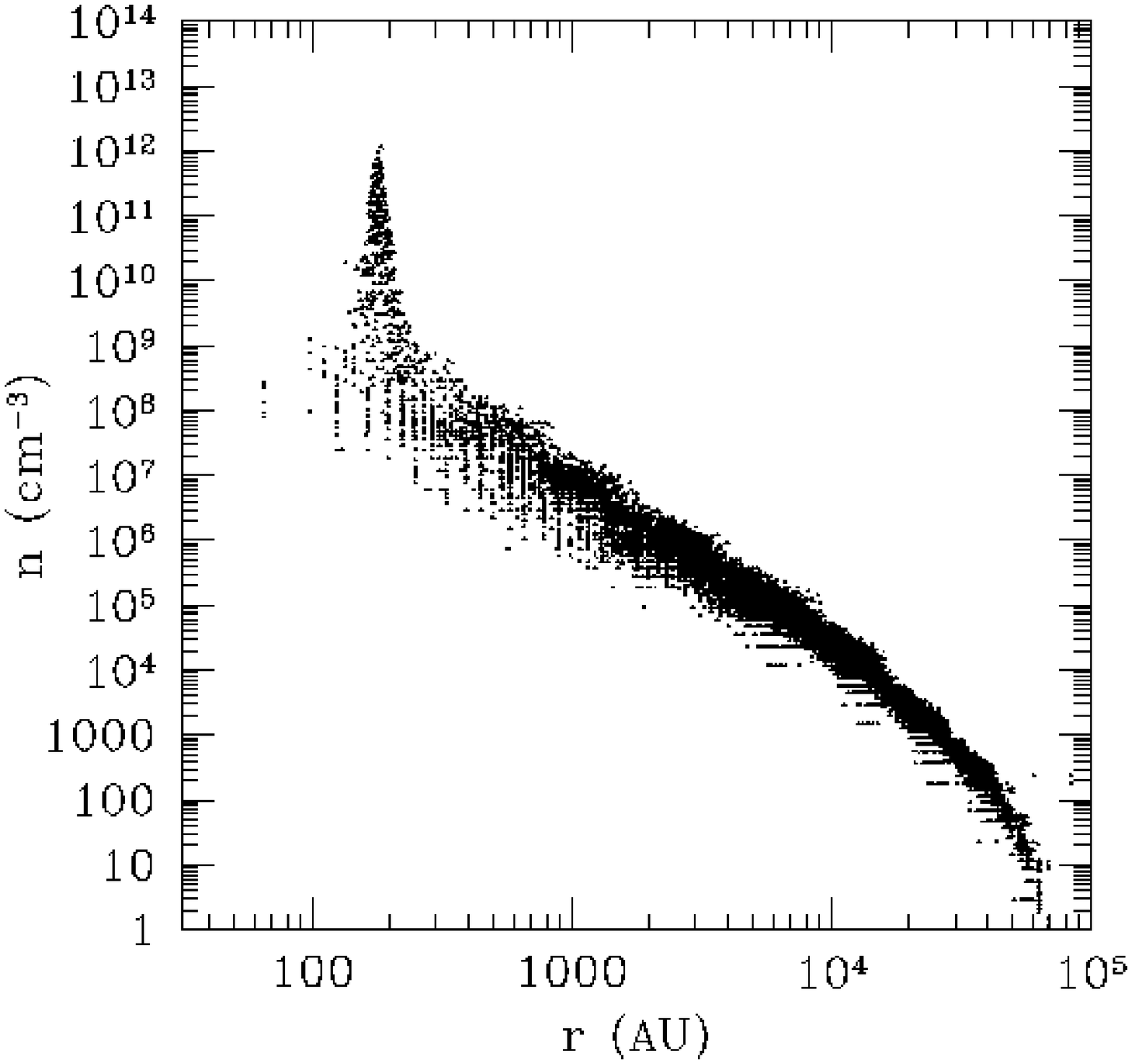}
\includegraphics[width=4.2cm]{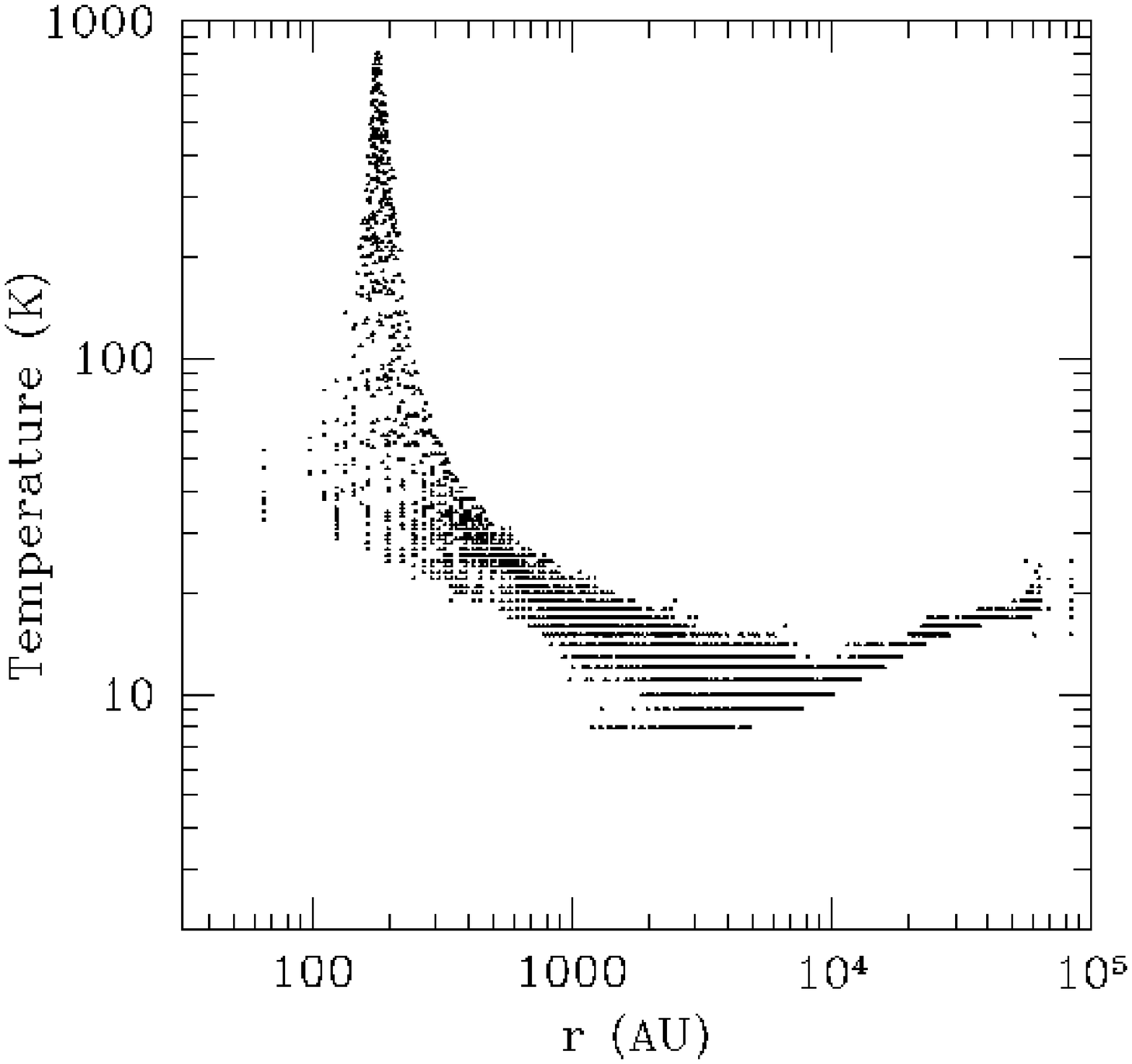}
}\vspace{-0.5cm}
\centerline{
\includegraphics[width=4.2cm]{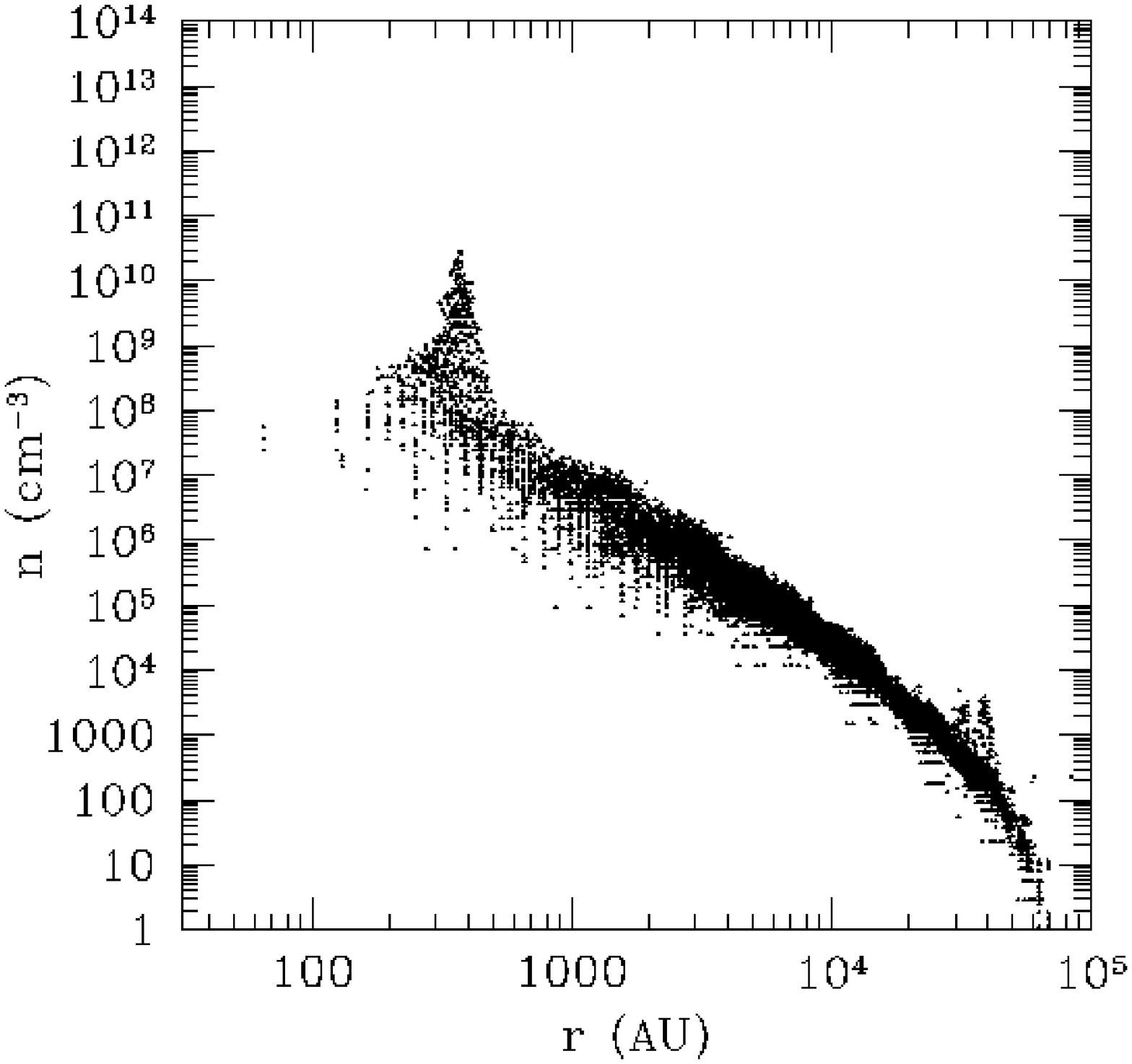}
\includegraphics[width=4.2cm]{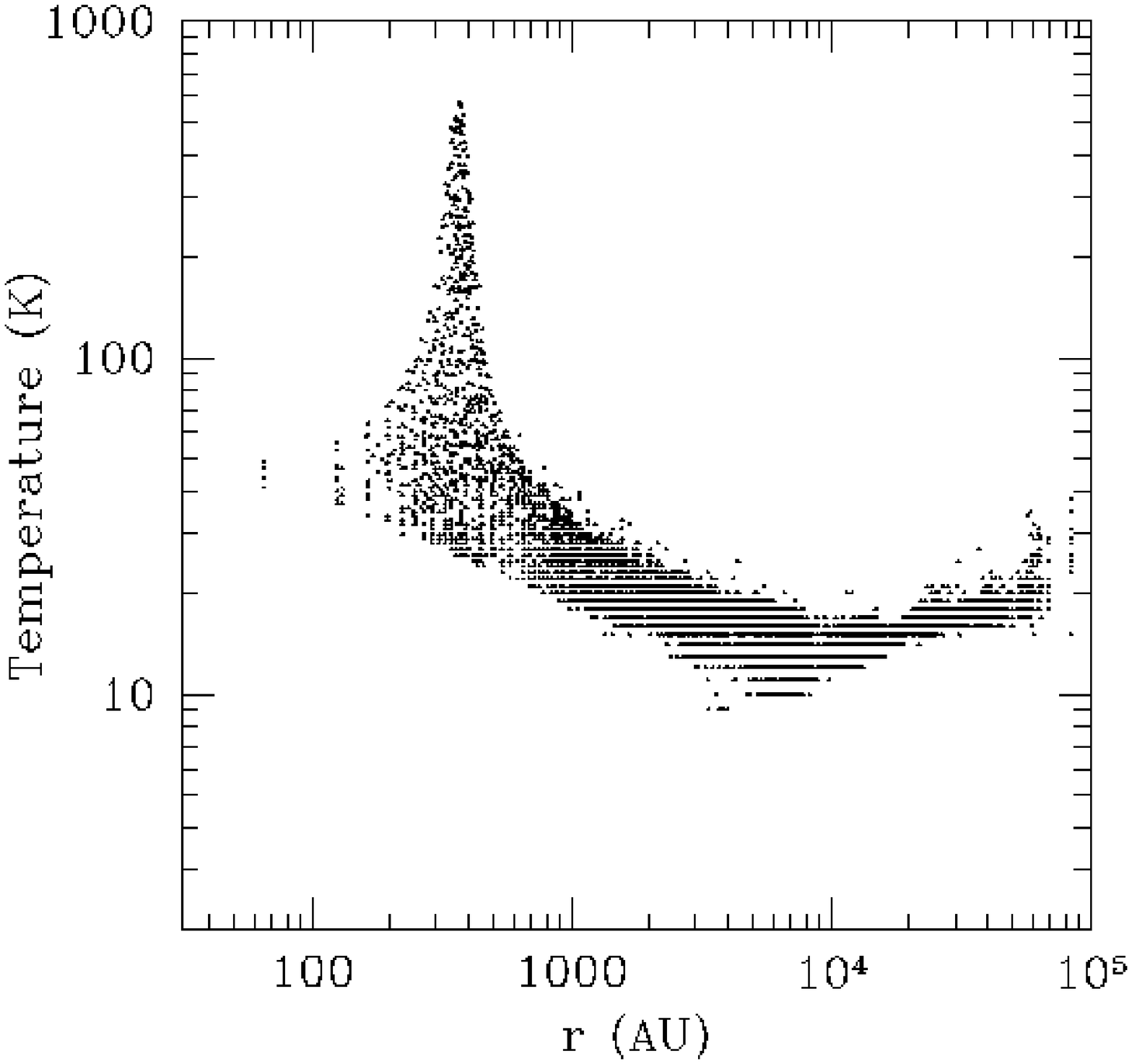}
}\vspace{-0.5cm}
\centerline{
\includegraphics[width=4.2cm]{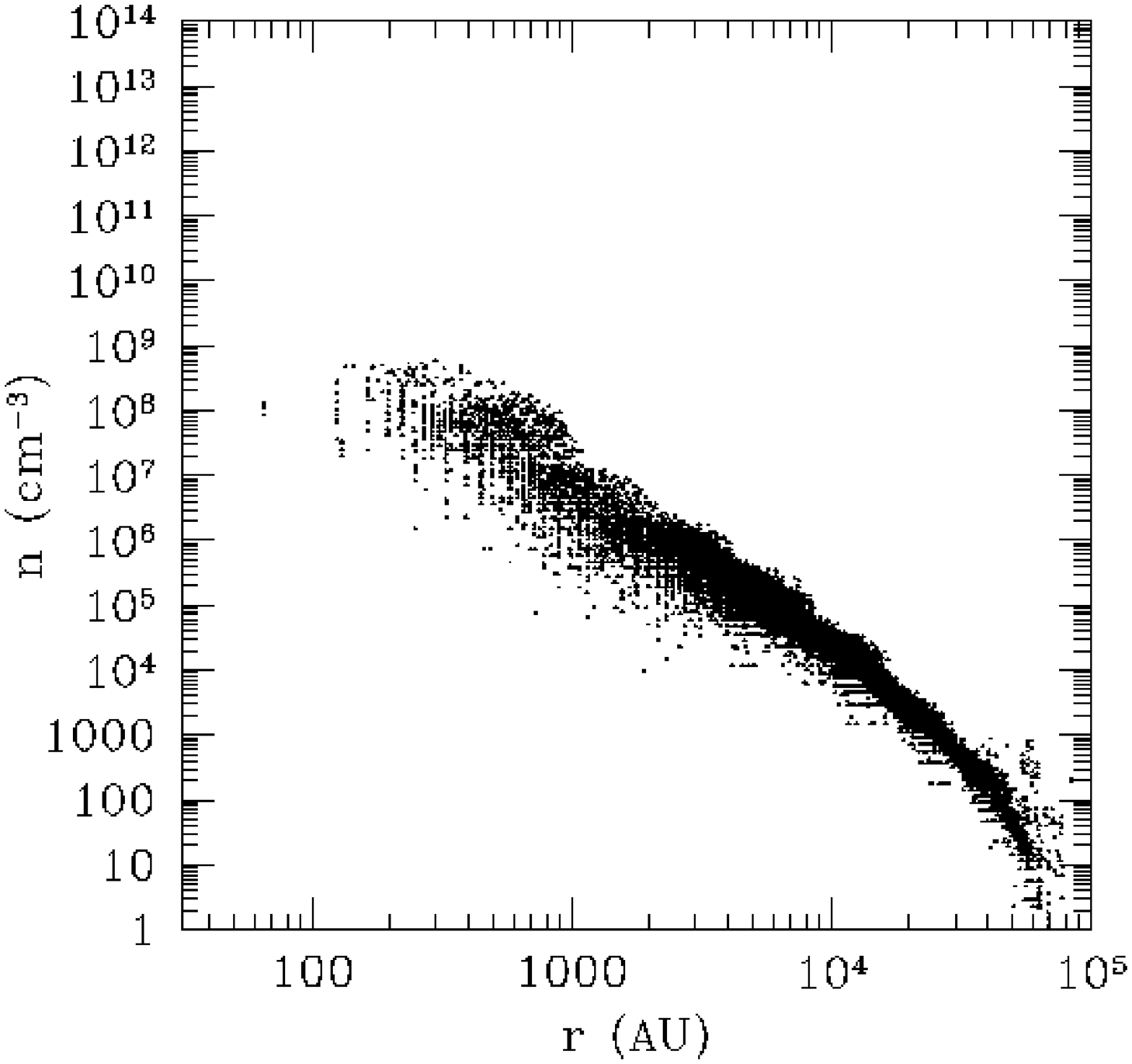}
\includegraphics[width=4.2cm]{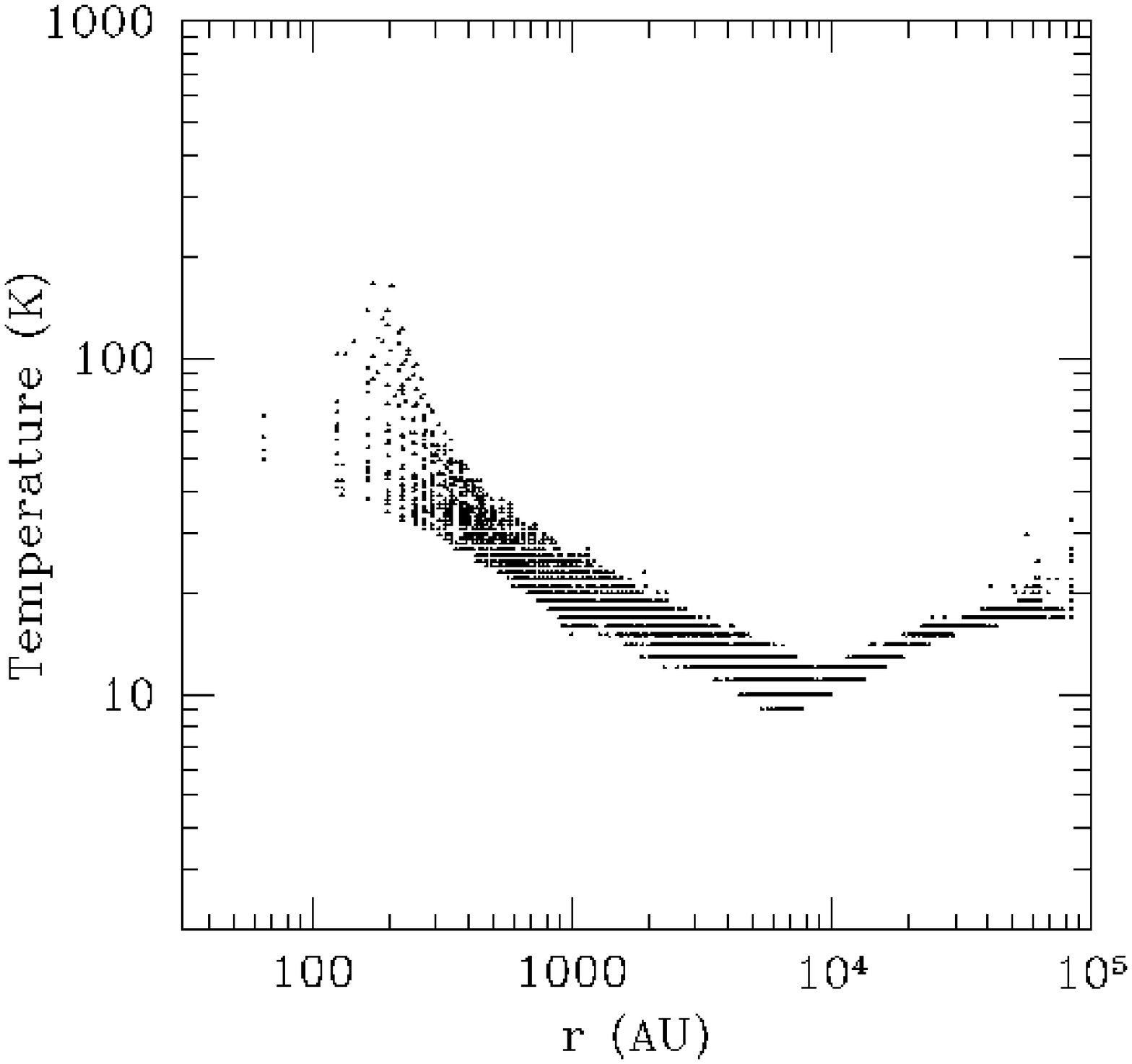}
}\vspace{-0.5cm}
\centerline{
\includegraphics[width=4.3cm]{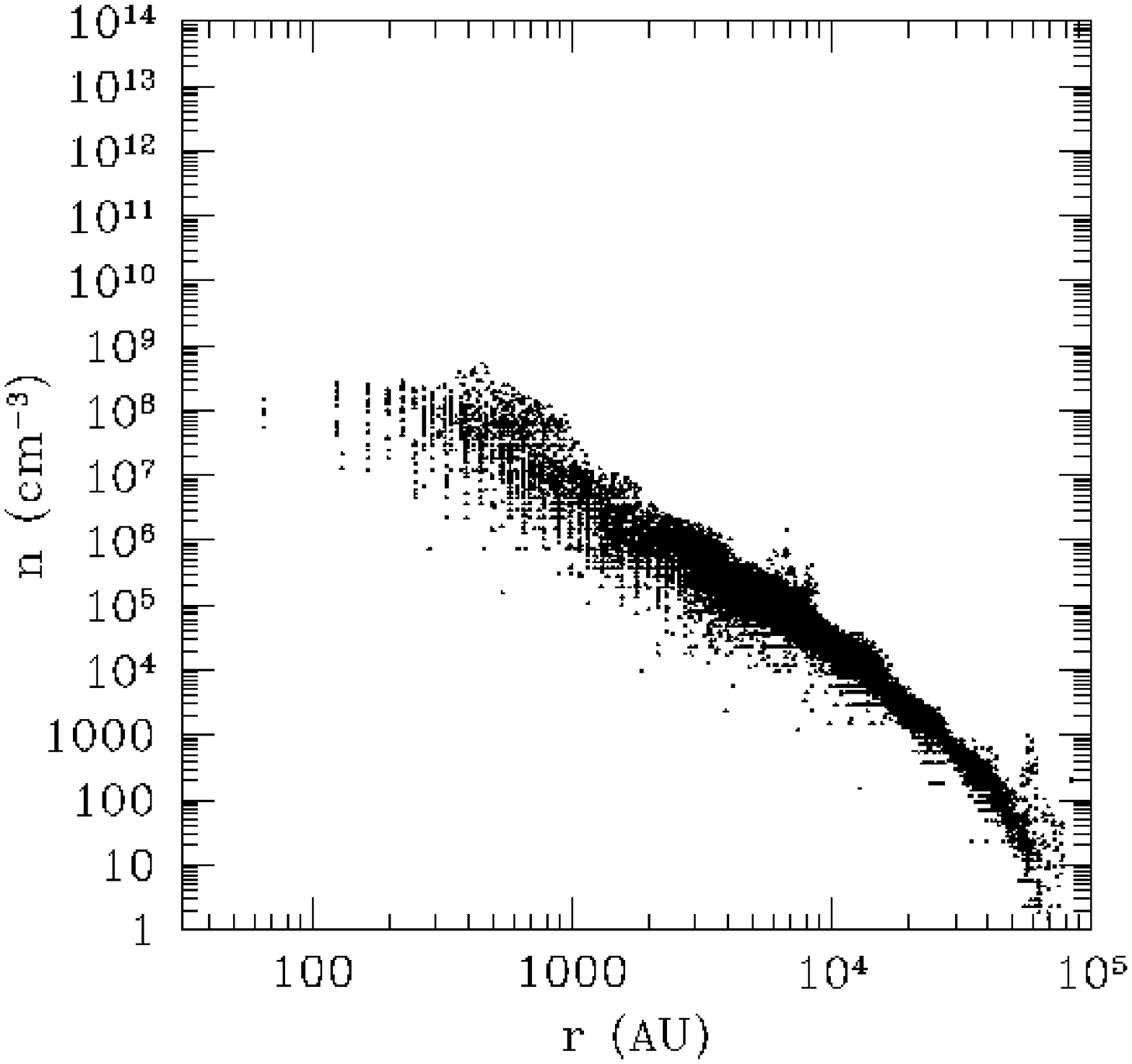}
\includegraphics[width=4.3cm]{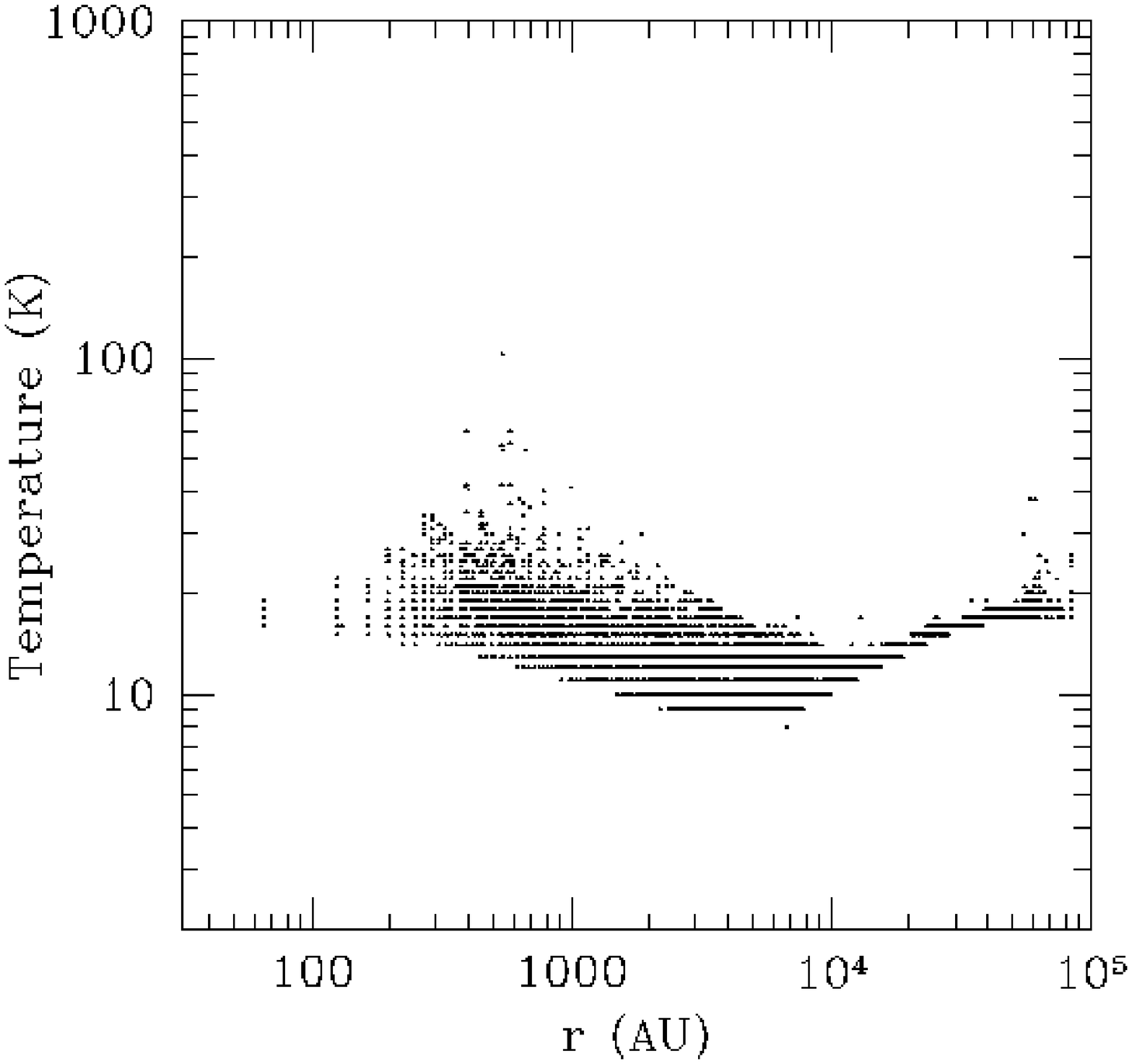}
}
\caption{Log-log plot of density (left column) and dust temperature (right column) 
versus distance from the centre of coordinates, in a collapsing core which forms a 
protostar; from top to bottom, time-frames {\texttt t0} to {\texttt t5} (see 
Table~\ref{tab:model.params}).}
\label{fig.log.dens.temp}
\end{figure}

In Fig.~\ref{fig.dens.temp} we present the density fields and dust-temperature 
fields for the 6 time-frames in Table~\ref{tab:model.params}. We plot the density 
and the temperature on 2 cuts through the computational domain, one parallel to 
the $x=0$ plane and one parallel to the $z=0$ plane. Each of these planes passes 
through the protostar, or, if there is no protostar (as in the first 2 time-frames), 
through the densest part of the core. The plots show only the central region of the 
core ($5000\,{\rm AU} \times 5000\,{\rm AU}$). Additionally, on 
Fig.~\ref{fig.log.dens.temp}, we plot the density and the temperature of each RT 
cell versus distance $r$ from the centre of coordinates, on a logarithmic scale, 
in order to depict better the regions very close to the protostar.

Our results for the temperature are broadly similar to those of previous 1D and 
2D studies of prestellar cores and Class 0 objects. Before the collapse, the 
precollapse prestellar core is quite cold ($10\;{\rm to}\,20\,{\rm K}$) and it becomes even 
colder ($5\;{\rm to}\,20\,{\rm K}$) as the collapse proceeds and the central regions 
become very dense and opaque. As soon as a protostar forms, the region around it 
becomes very hot (up to the dust destruction temperature), but the temperature still 
drops below $\sim\!20\,{\rm K}$ beyond a few $1000\,{\rm AU}$ from the protostar, 
because of the high optical depth in the dense accretion flow onto the protostar. 
The luminosity of the protostar is dominated by the contribution from accretion. 
Initially the accretion luminosity increases due to the increasing mass of the 
protostar, but then it falls as the accretion rate declines.

\subsection{SEDs of prestellar cores and young protostars}

\begin{figure*}
\centerline{
\includegraphics[width=5.8cm]{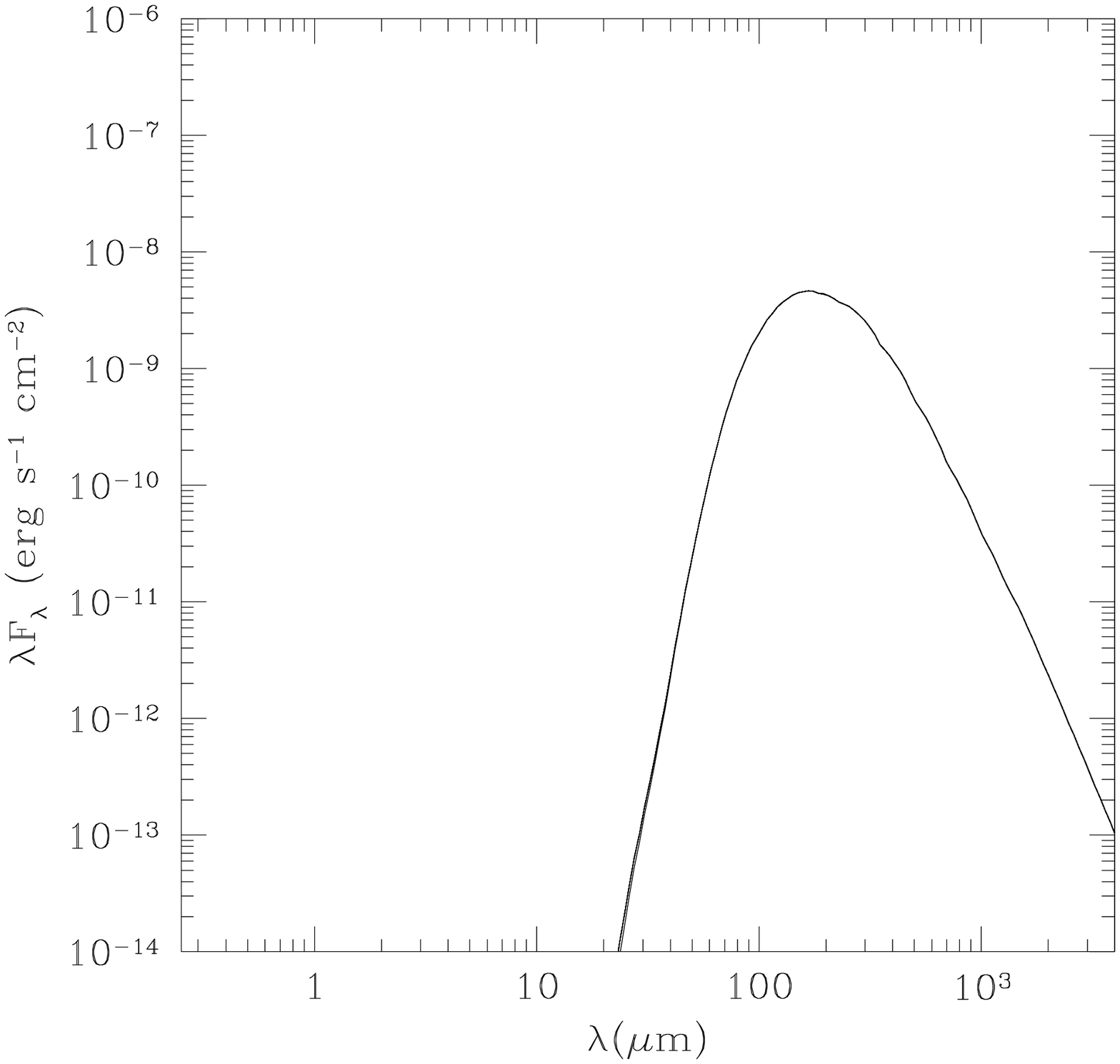}
\includegraphics[width=5.8cm]{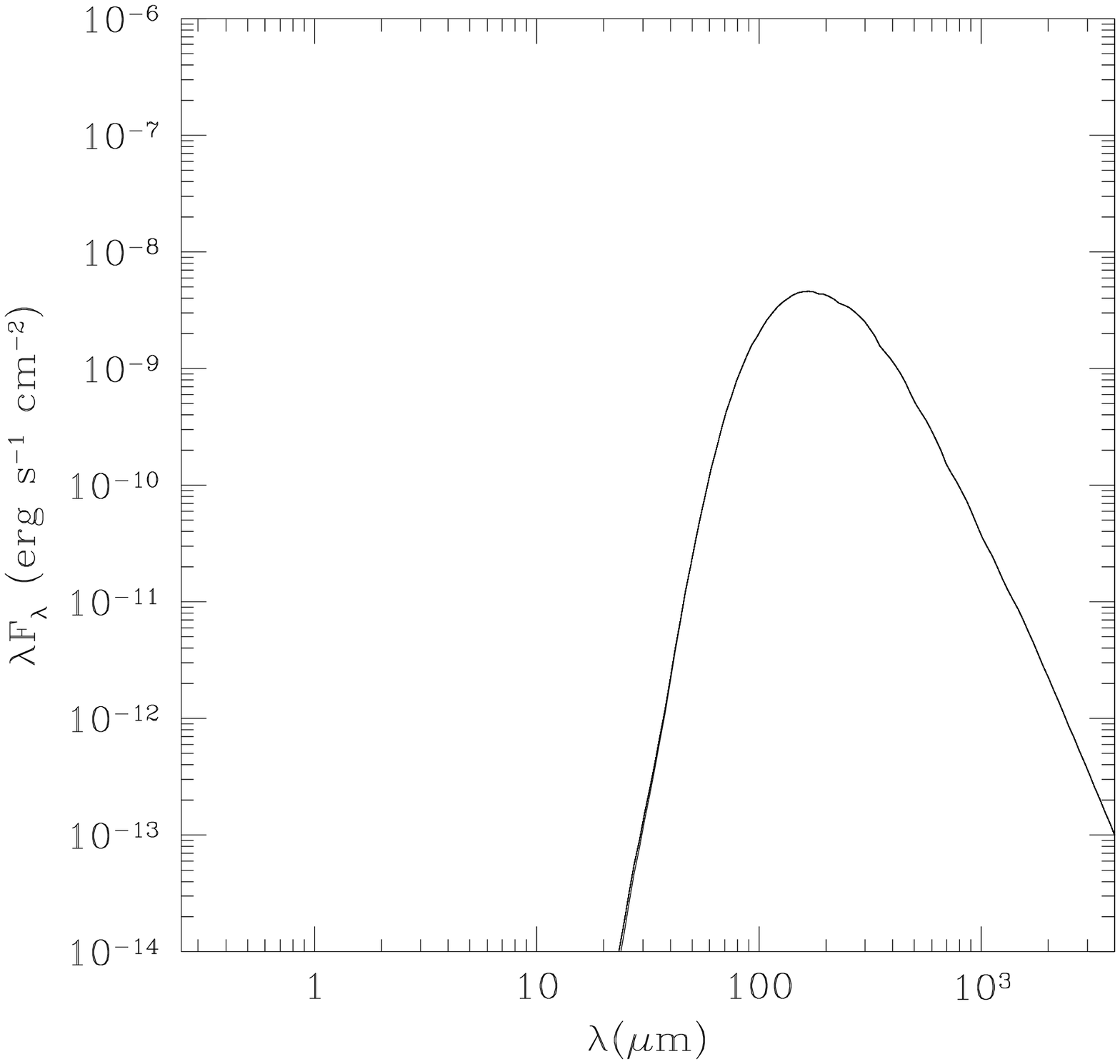}
\includegraphics[width=5.8cm]{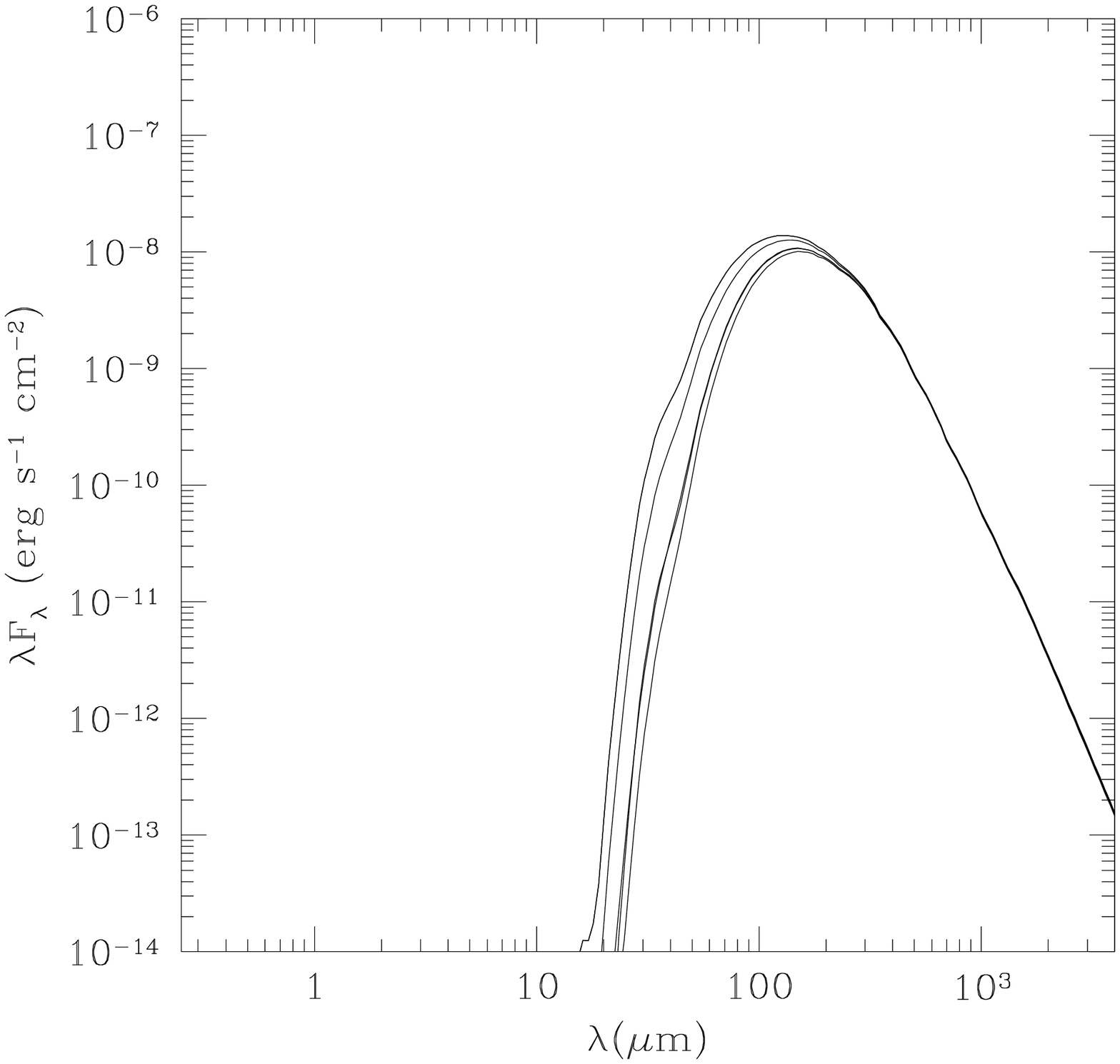}}
\centerline{
\includegraphics[width=5.8cm]{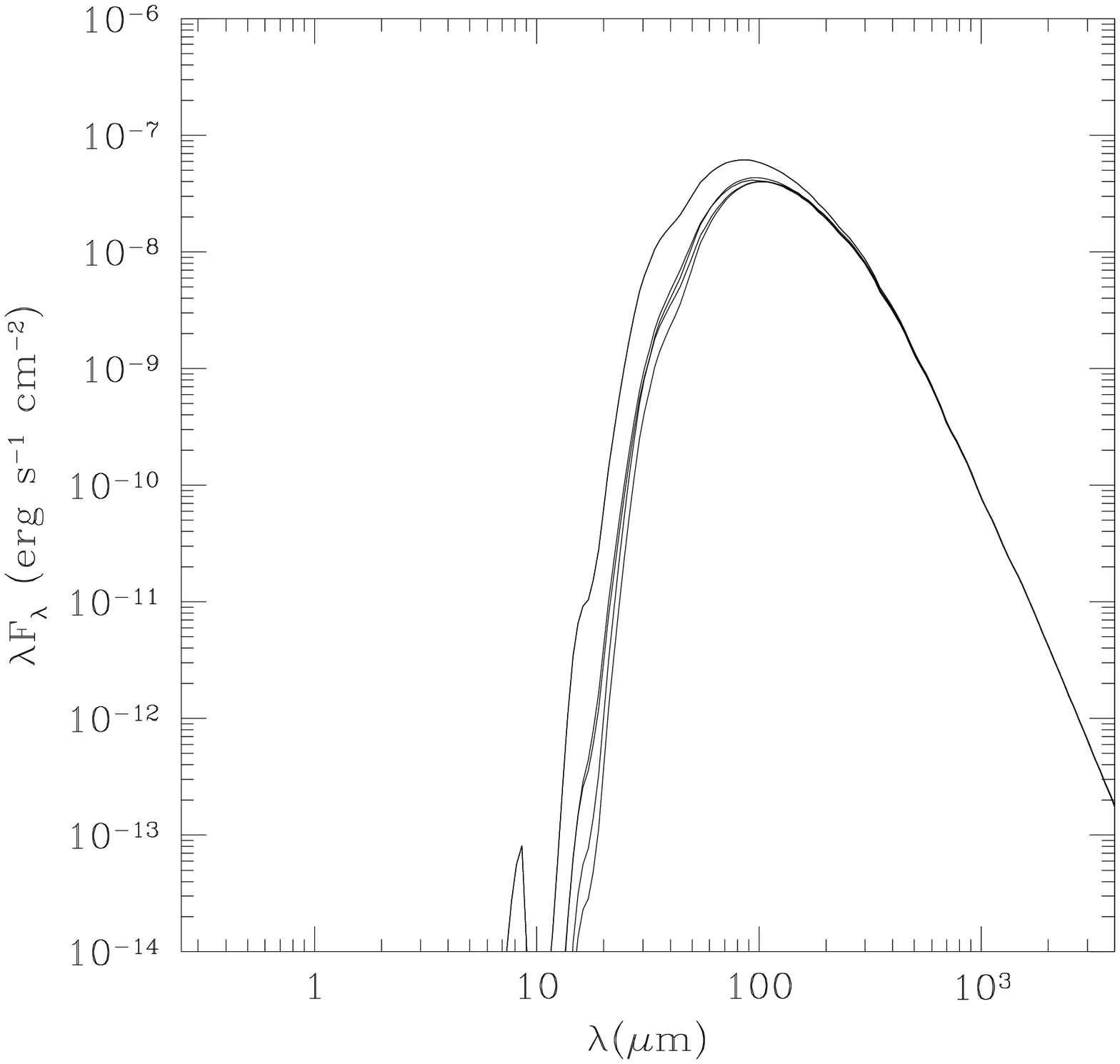}
\includegraphics[width=5.8cm]{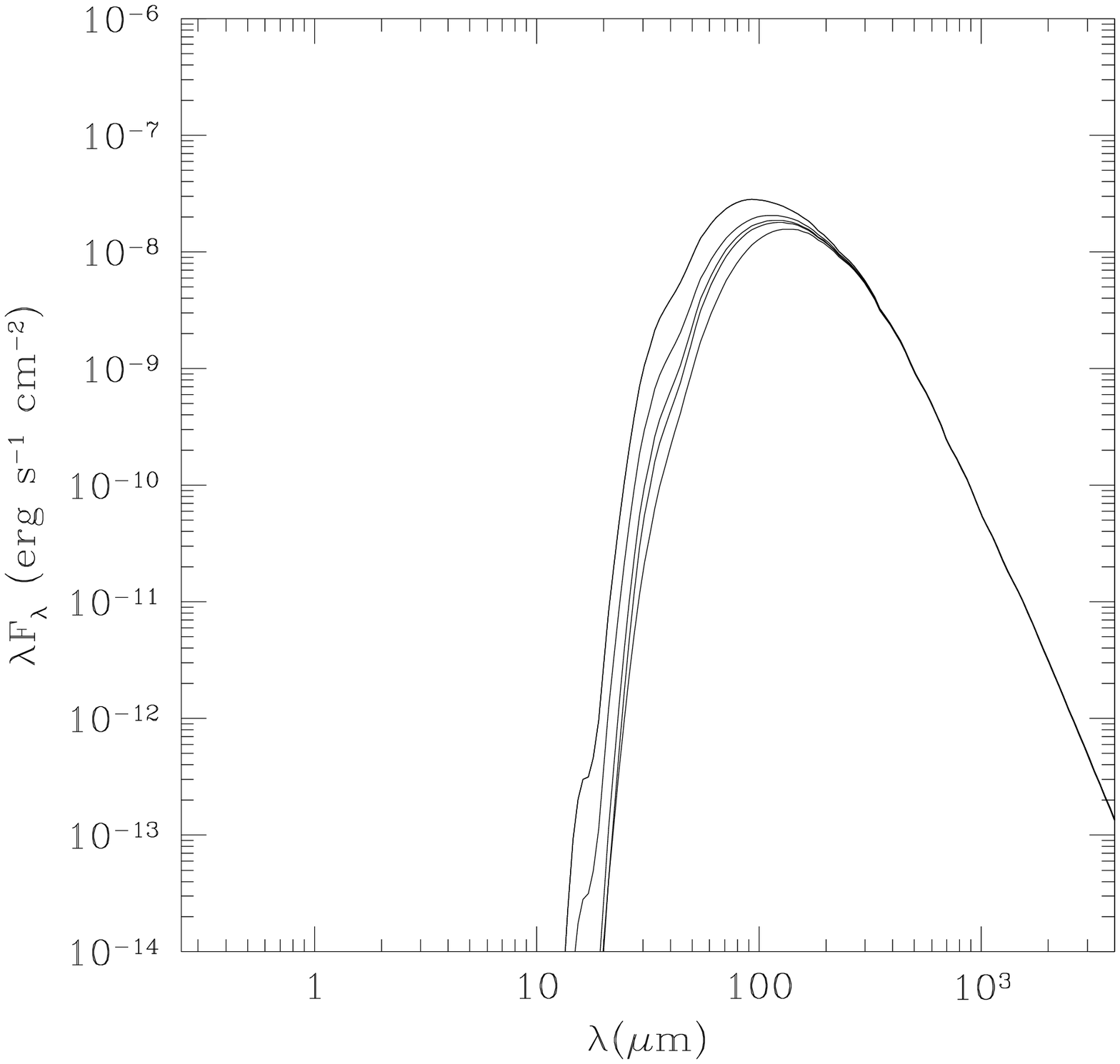}
\includegraphics[width=5.8cm]{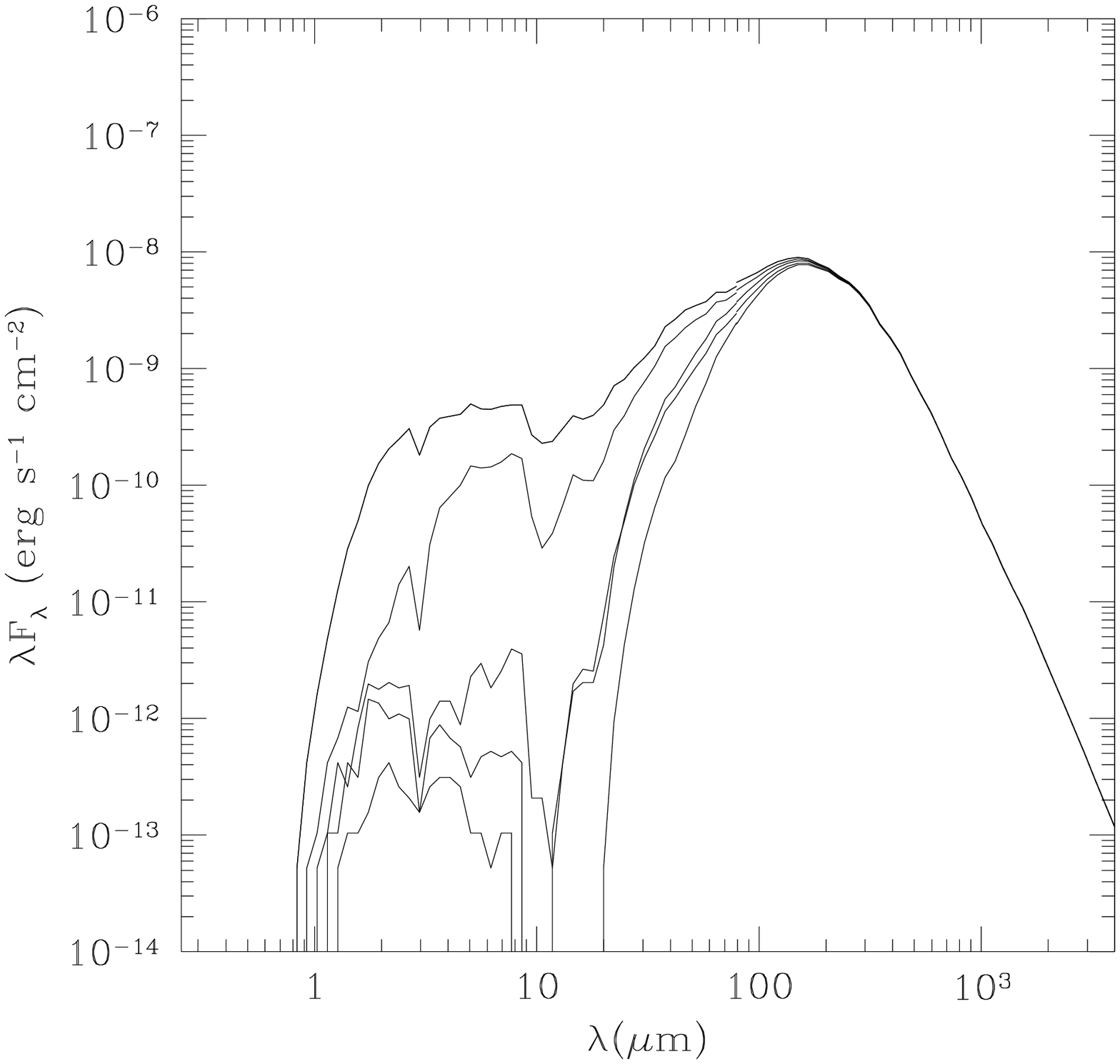}}
\caption{SEDs of the 6 time-frames in Table~\ref{tab:model.params}, from 3 polar 
angles ($\theta=0\degr$, $45\degr$, $90\degr$) and 2 azimuthal angles ($\phi=0\degr$, 
$90\degr$); in several cases the SEDs from different viewing angles overlap. {\bf (a)} 
{\texttt t0} - {\it precollapse prestellar core.} The SED peaks at $\sim 190\,\micron$, which 
implies an effective temperature $T_{_{\rm EFF}} \sim 13\,{\rm K}$. {\bf (b)} {\texttt t1} 
- {\it collapsing prestellar core.} The SED appears the same as in (a) although the 
collapse has started and the cloud is denser and colder at its centre. {\bf (c)} 
{\texttt t2} - {\it Class 0 object.} The emission of the system peaks at $\lambda \sim 
100\;{\rm to}\,135\,\micron$ (depending on the observer's  viewing angle), which implies 
$T_{_{\rm EFF}} \sim 18\,{\rm to}\,25\,{\rm K}$. The total luminosity of the system is 
higher than in (b). {\bf (d)} {\texttt t3} - {\it Class 0 object.} Similar to (c), but 
the luminosity is even higher, and the SED peaks at $\lambda \sim 80\;{\rm to}\,100\,\micron$, 
implying $T_{_{\rm EFF}} \sim 25\;{\rm to}\,31\,{\rm K}$. {\bf (e)} {\texttt t4} - {\it Class 
0 object.} Similar to (d), but the luminosity has started to decrease due to the falling 
accretion rate, and the SED peaks at $\lambda \sim 90\;{\rm to}\,135\,\micron$, implying 
$T_{_{\rm EFF}} \sim 18\;{\rm to}\,28\,{\rm K}$. {\bf (f)} {\texttt t5} - {\it Class 0 
object.} The protostar has moved out of the central region by $\sim 100\,{\rm AU}$, and 
the attenuated stellar emission can be seen at short wavelengths ($1\;{\rm to}\,50\,\micron$). 
The peak of the cloud emission is at $\lambda \sim 150\,\micron$, implying $T_{_{\rm EFF}}
\sim 17\,{\rm K}$.}
\label{fig.seds}
\end{figure*}

The SEDs of the 6 time-frames in Table~\ref{tab:model.params} are presented in
Fig.~\ref{fig.seds}. These SEDs have been calculated assuming that the core is 
at $140\,{\rm pc}$. SEDs are plotted for 6 different viewing angles, i.e. 3 polar 
angles ($\theta=0\degr$, $45\degr$, $90\degr$) and 2 azimuthal angles 
($\phi=0\degr$, $90\degr$). Hence on each graph there are 6 curves. However, in 
several cases the SED is so weakly dependent on viewing angle that the curves 
overlap. We plot only the thermal emission from the core, i.e. we neglect both 
the long-wavelength background radiation that just passes through the cloud, and 
the short-wavelength radiation that is scattered by the outer layers of the cloud. 

The effective temperature of the core, as inferred from the peak of the SED, rises 
and falls with the accretion luminosity of the protostar. For a prestellar core, 
the SED peaks at $\sim 190\,\micron$, implying $T_{_{\rm EFF}} \sim 13\,{\rm K}$. 
Once the protostar has formed and inputs energy into the system, the peak moves 
steadily to shorter wavelengths, reaching $\sim 80\,\micron$ ($T_{_{\rm EFF}} \sim 
31\,{\rm K}$) as the accretion luminosity reaches its maximum, and then moving back 
to longer wavelengths again as the accretion luminosity declines. By the final frame 
it has reached $\sim 150\,\micron$ ($T_{_{\rm EFF}} \sim 17\,{\rm K}$).

The peak of the SED of a prestellar core is independent of viewing angle, since 
the core is optically thin to the radiation it emits. In contrast, the peak of the 
SED of a Class 0 object does depend on viewing angle, albeit weakly, because of the 
presence of an optically thick disc around the protostar. However, even allowing for 
variations in the viewing angle, the SED of a Class 0 object does not peak at the 
wavelengths characteristic of prestellar cores ($\sim 190\,\micron$).

\subsection{Is a young Class 0 observable in the NIR?}

Our model predicts that a young protostar embedded in a core 
is not observable in the NIR, {\it unless} the protostar is displaced from 
the central high-density region. This contradicts the recent results of 
Whitney et al. (2003b) and Young et al.~(2004). We attribute the difference 
between our model and that of Young et al. to the density they use for the 
region around the protostar ($\sim 10^5\,{\rm cm}^{-3}$), which may be too 
low for a Class 0 objects and more appropriate for a prestellar core. 
Whitney et al. use a higher density in the central region ($\sim 10^{10}\,
{\rm cm}^{-3}$), but it is still more than one order of magnitude lower 
than the density produced by our SPH simulation ($\sim 3 \times 10^{12}\,
{\rm cm}^{-3}$). Additionally, they use different dust properties for the 
different regions in their configuration (cloud, disc mid-plane, disc atmosphere, 
outflow); in particular, the opacity of their dust in the disc midplane is smaller 
than the opacity in our model, at wavelengths shorter than $10\,\micron$. There 
are also some differences in the properties of the discs used in the different 
models, but these are less relevant to the escape of NIR radiation.

The models reported by Shirley et al. (2002) use central densities and dust 
opacities similar to our models, but they confine their study to the FIR/mm 
region of the spectrum, so a comparison is not possible.

Ultimately, the density distribution within $500\,{\rm AU}$ of the centre of 
a core or protostar is not well constrained by observations, and therefore 
radiative transfer calculations have to rely on theoretical models of core 
collapse. Dust opacities are also poorly constrained (e.g. Bianchi et al. 
2003). Comparison of our model with those of Whitney et al. (2003b) and Young 
et al. (2004) suggests that the presence or absence of NIR emission from a 
confirmed Class 0 object might be used to constrain the density profile within 
a few $100\,{\rm AU}$ of the protostar and/or the opacity of the dust in the 
same region.

\begin{figure*}
\centerline{
\includegraphics[width=6.65cm]{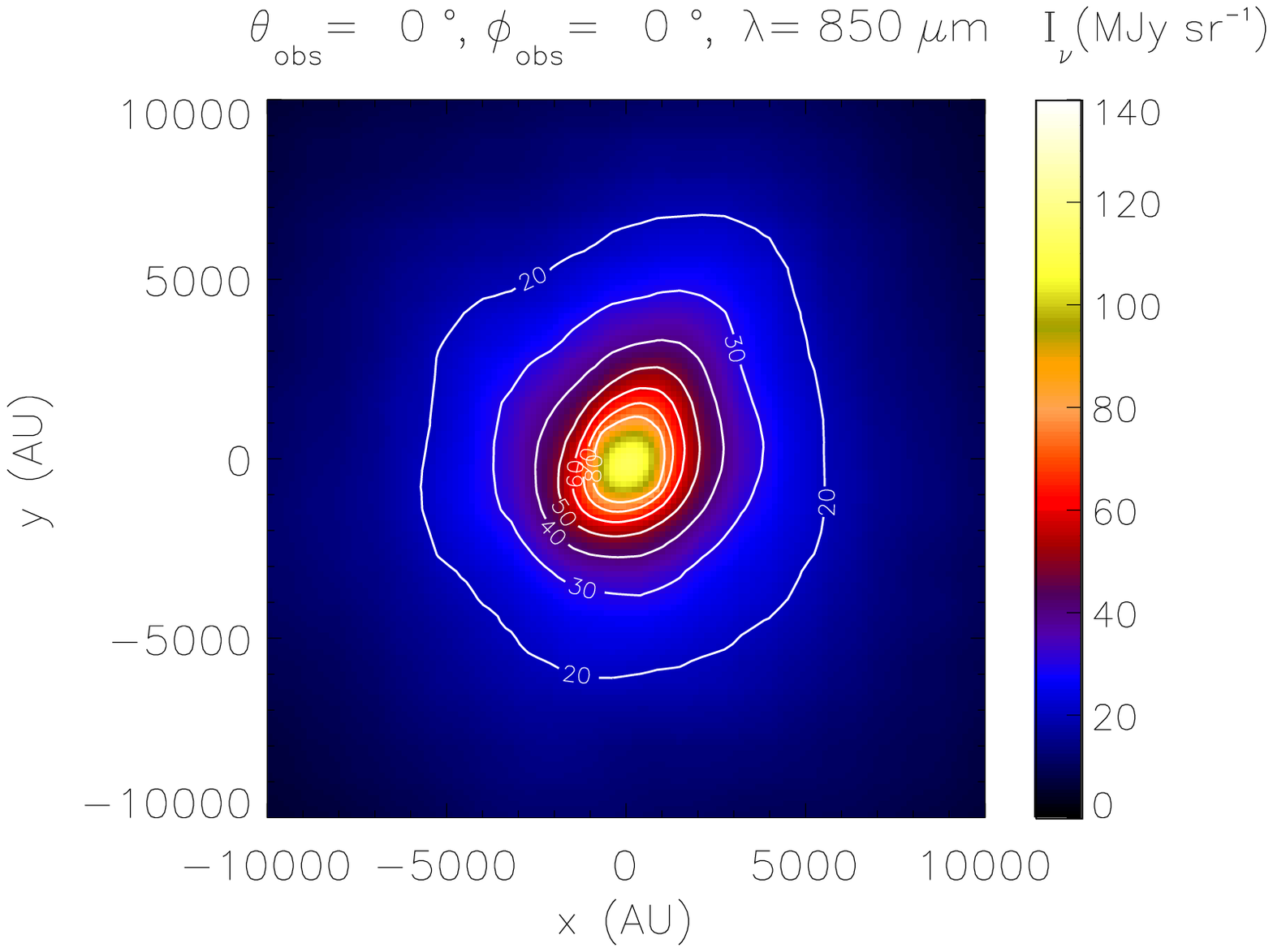}\hspace{-.3cm}
\includegraphics[width=6.65cm]{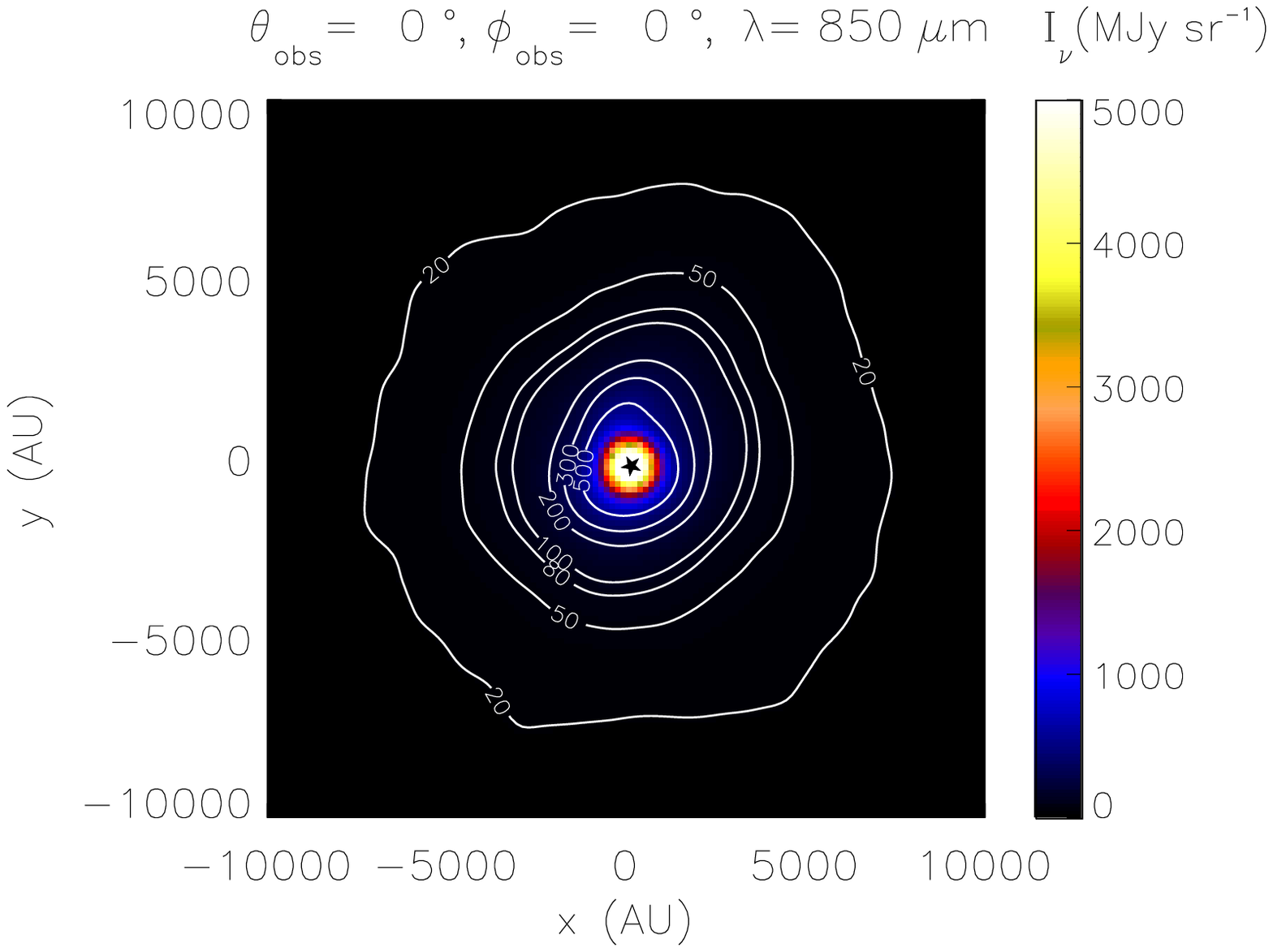}\hspace{-.3cm}
\includegraphics[width=6.65cm]{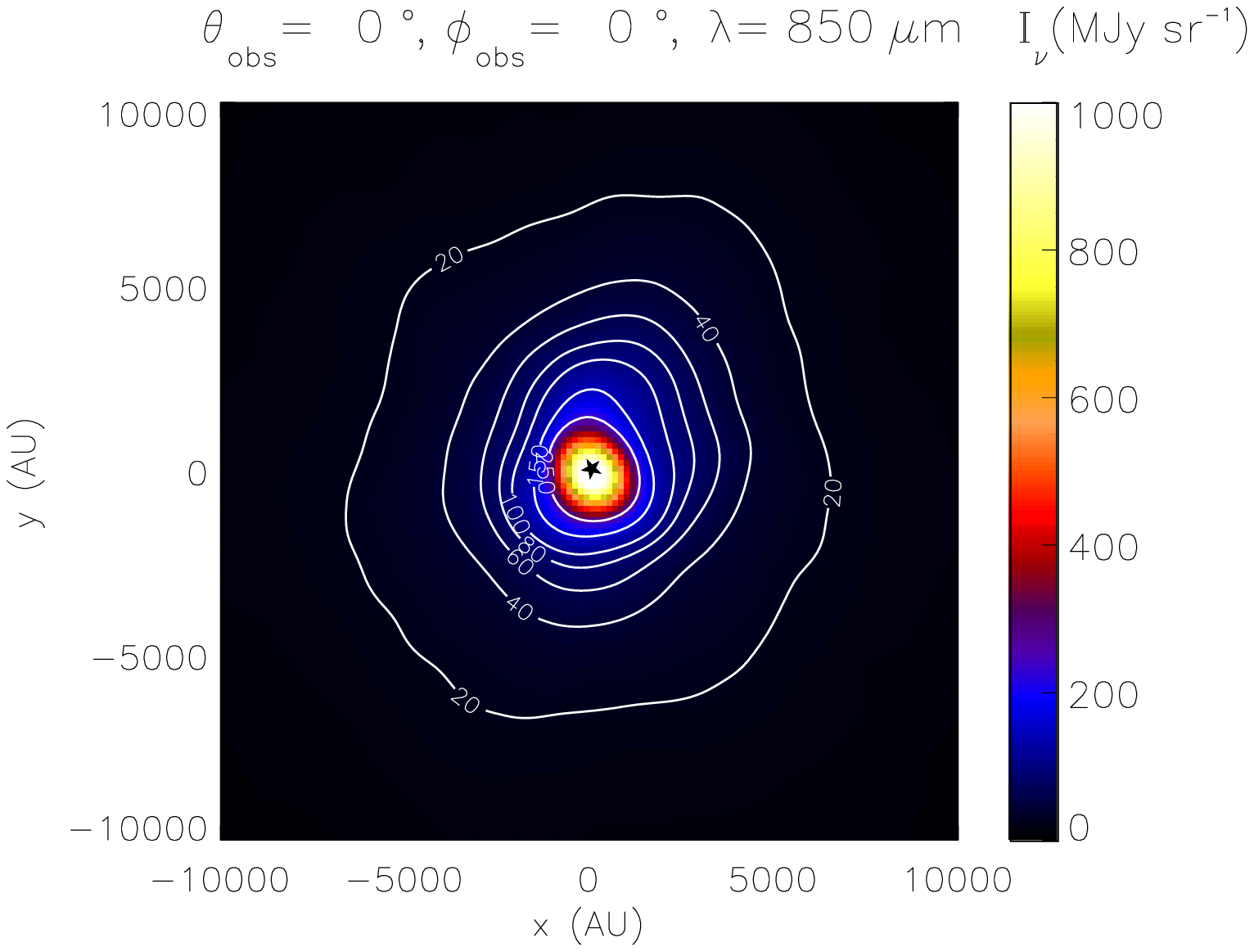}}
\centerline{
\includegraphics[width=6.65cm]{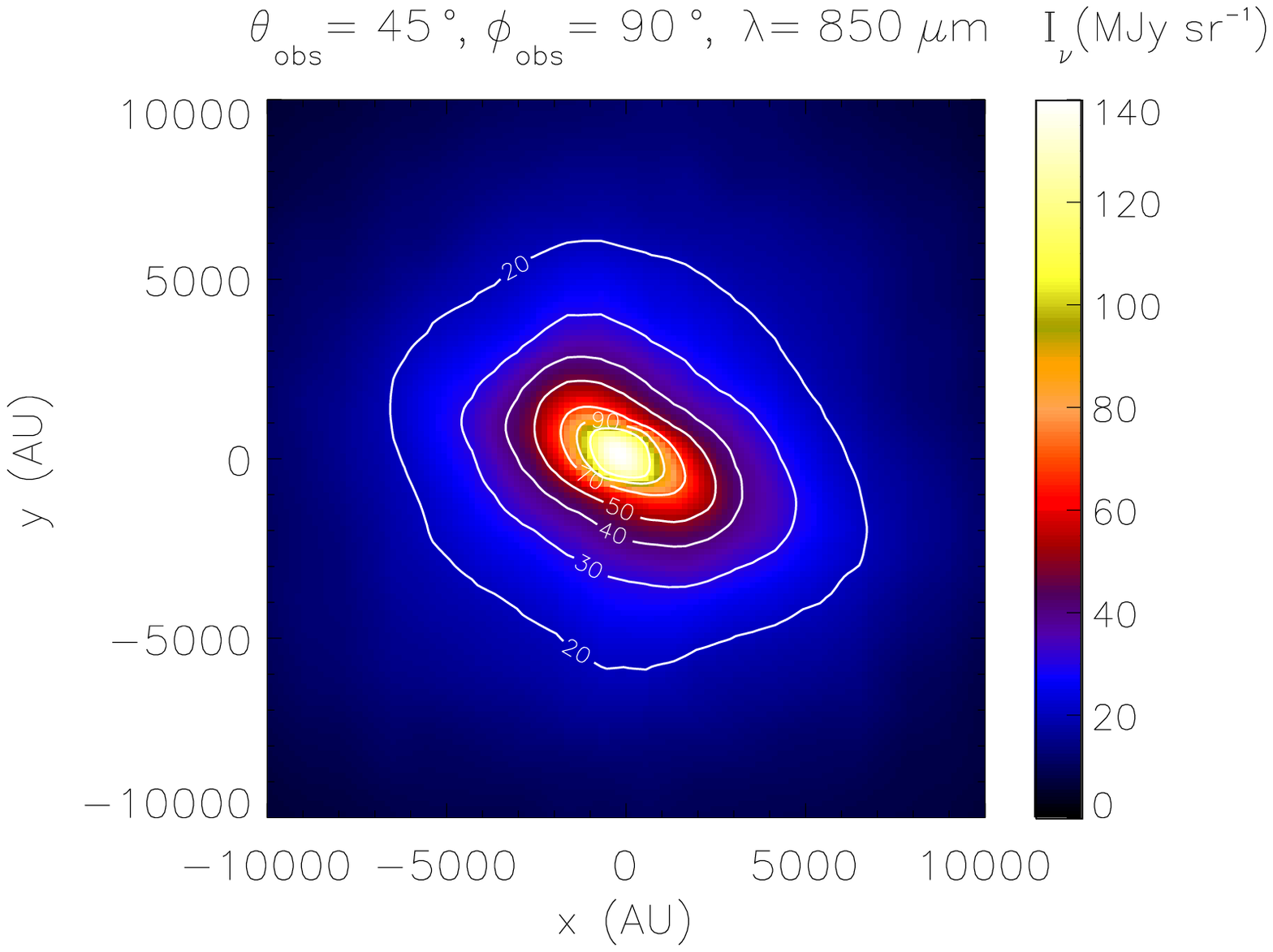}\hspace{-.3cm}
\includegraphics[width=6.65cm]{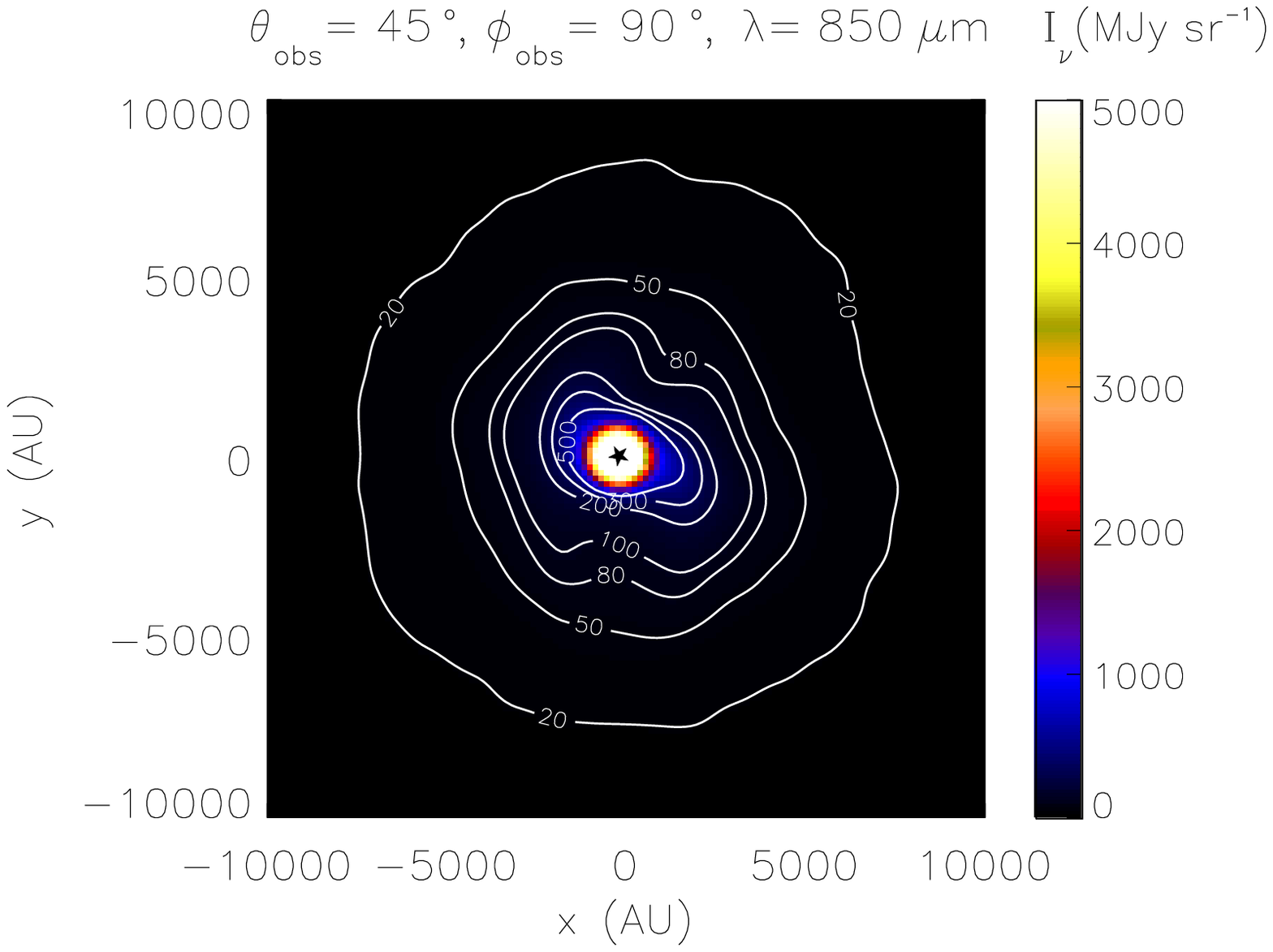}\hspace{-.3cm}
\includegraphics[width=6.65cm]{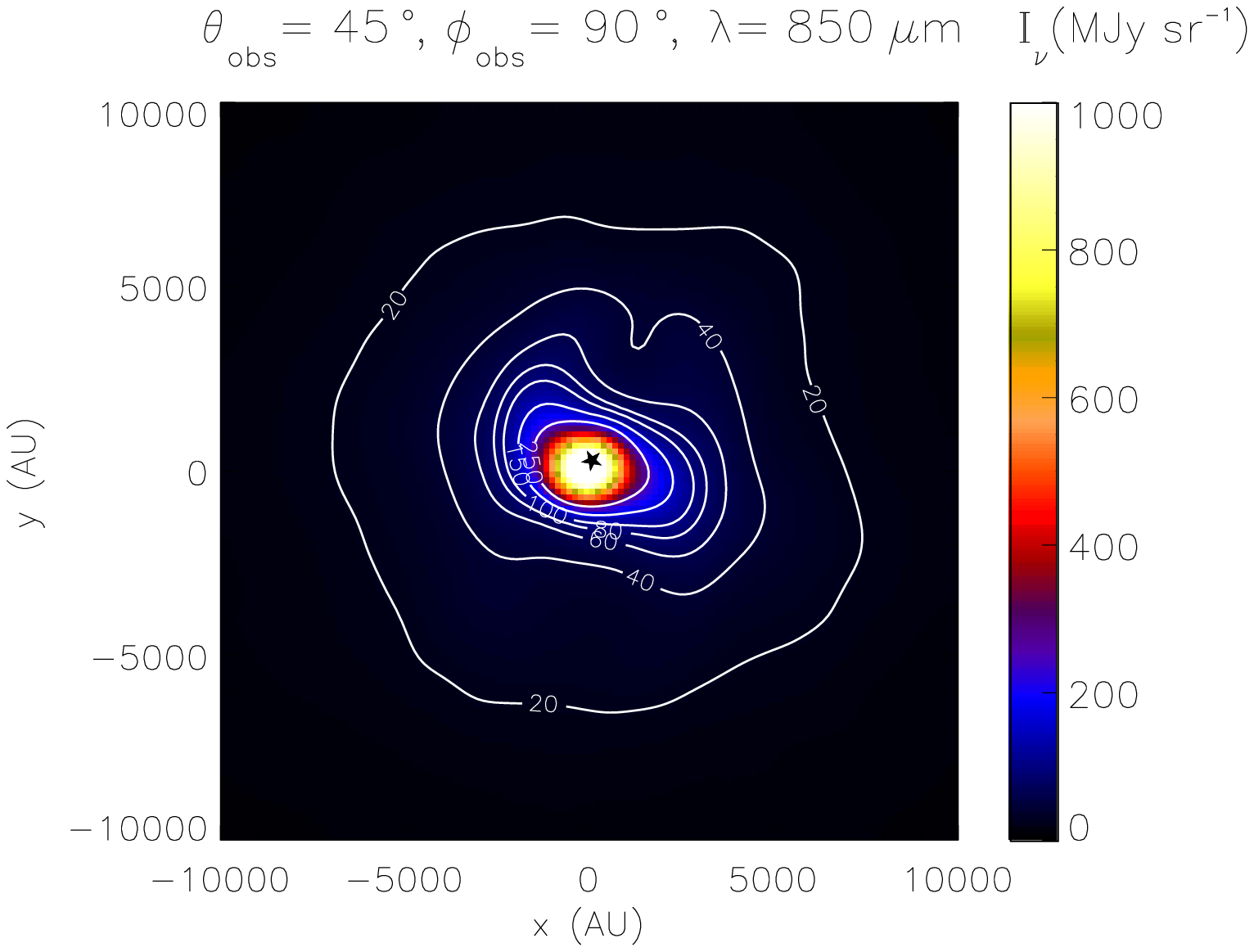}}
\centerline{
\includegraphics[width=6.65cm]{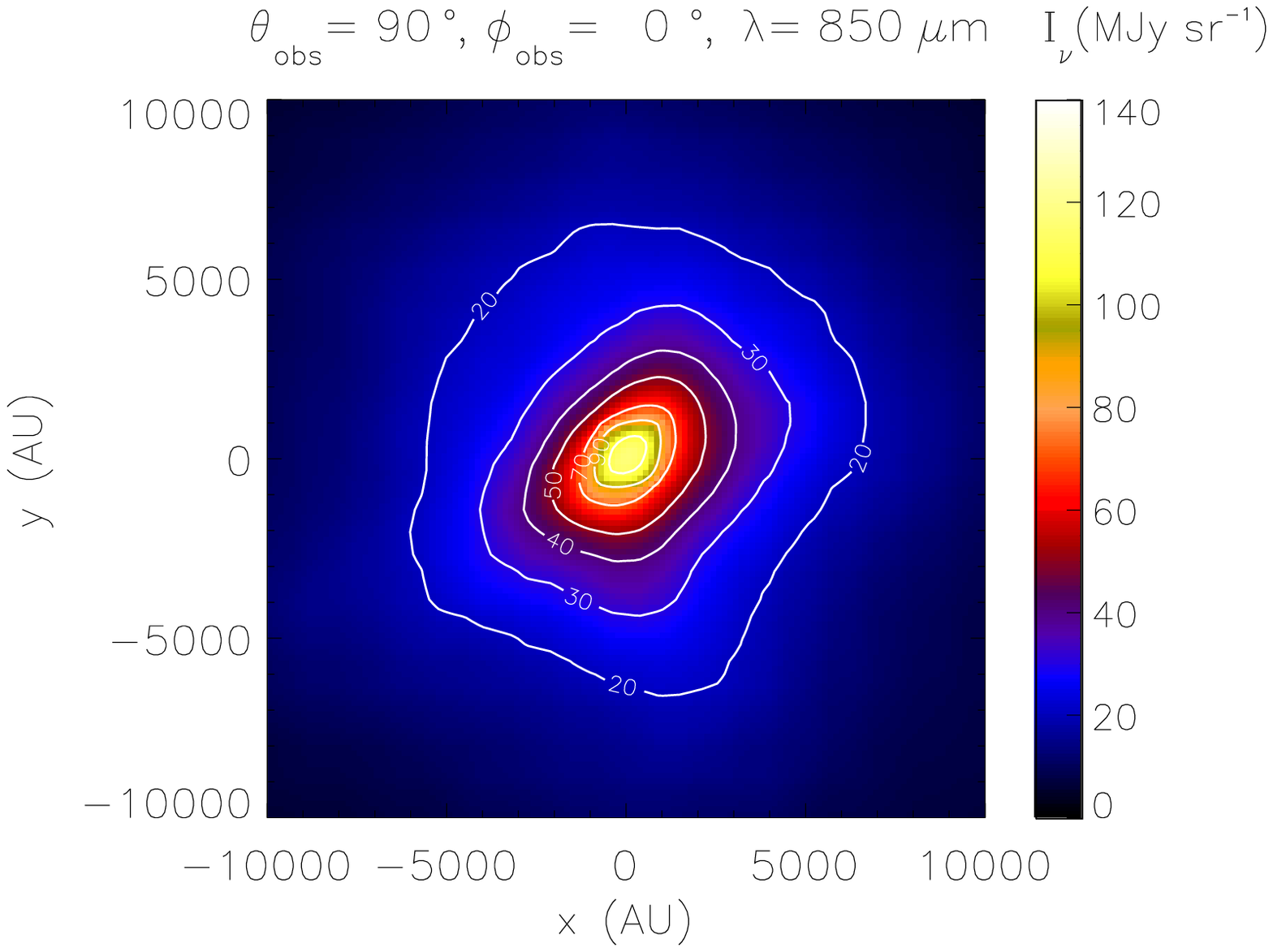}\hspace{-.3cm}
\includegraphics[width=6.65cm]{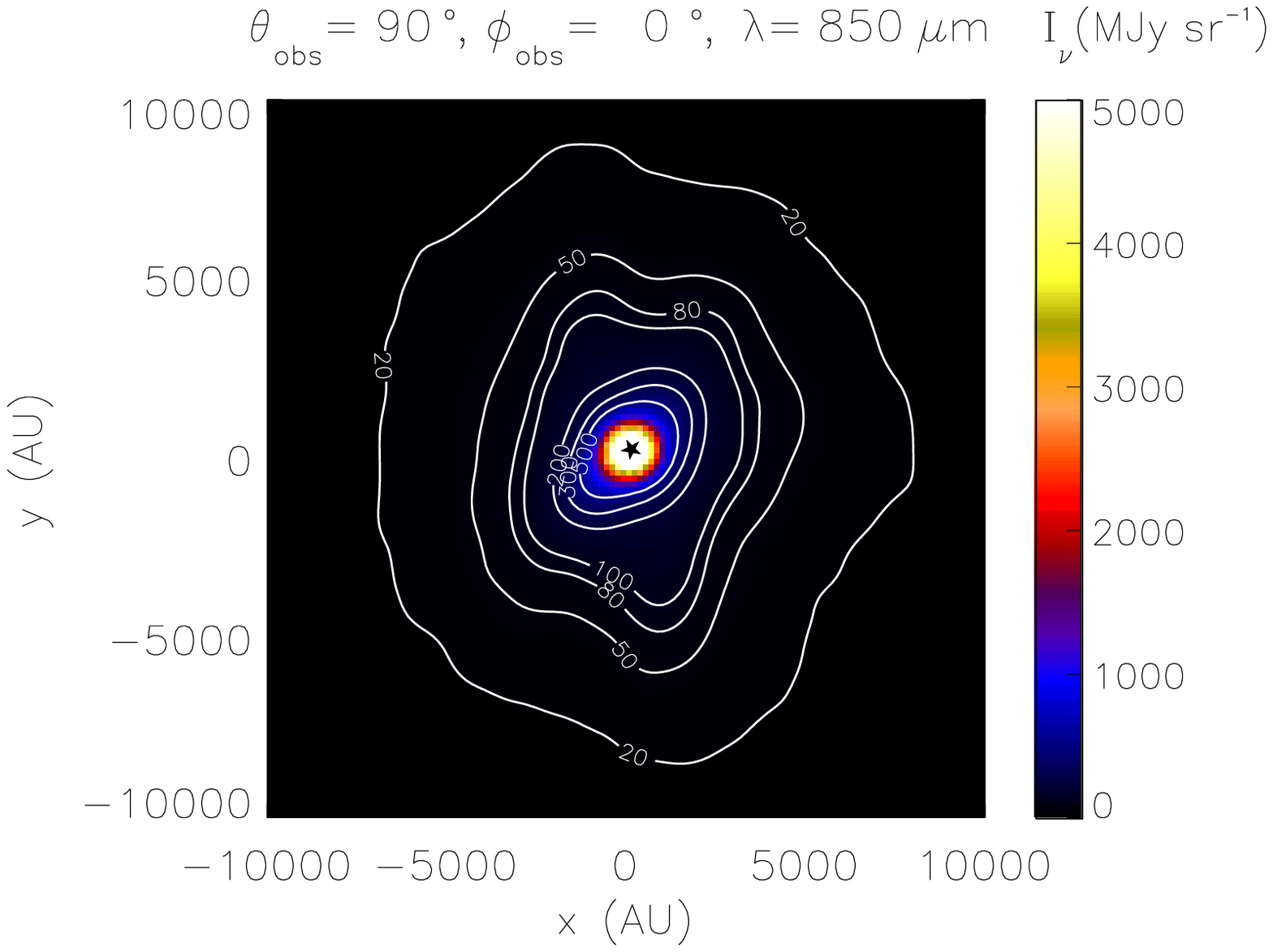}\hspace{-.3cm}
\includegraphics[width=6.65cm]{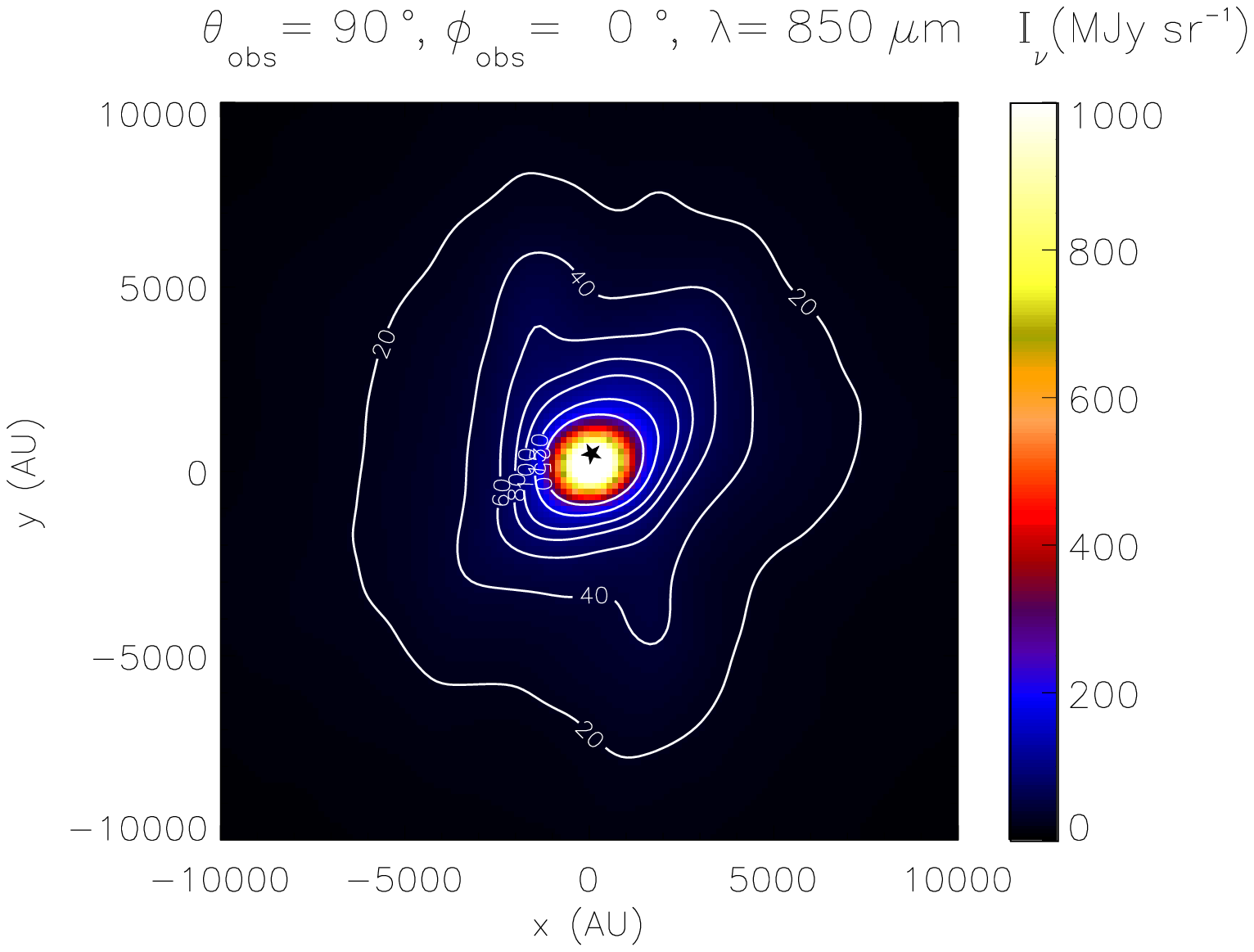}}
\caption{850 $\micron$ isophotal maps of 3 different time-frames (first column: 
{\texttt t1} - {\it collapsing prestellar core};$\;$ second column: {\texttt t3} - 
{\it Class 0 object};$\;$ third column: {\texttt t5} - {\it Class 0 object}), and 
from three different viewing angles (first row: $\theta_{\rm obs} = 0\degr,\;
\phi_{\rm obs} = 0\degr$;$\;$ second row: $\theta_{\rm obs} =45\degr,\;
\phi_{\rm obs} = 90\degr$;$\;$ third row: $\theta_{\rm obs} = 90\degr,\;
\phi_{\rm obs} = 0\degr$). Cores with protostars (second and third columns) 
are more centrally condensed than prestellar cores (first column).
We note that the axes $(x,y)$ refer to the plane of sky 
as seen by the observer.
}
\label{fig.isomaps}
\centerline{
\includegraphics[width=6.65cm]{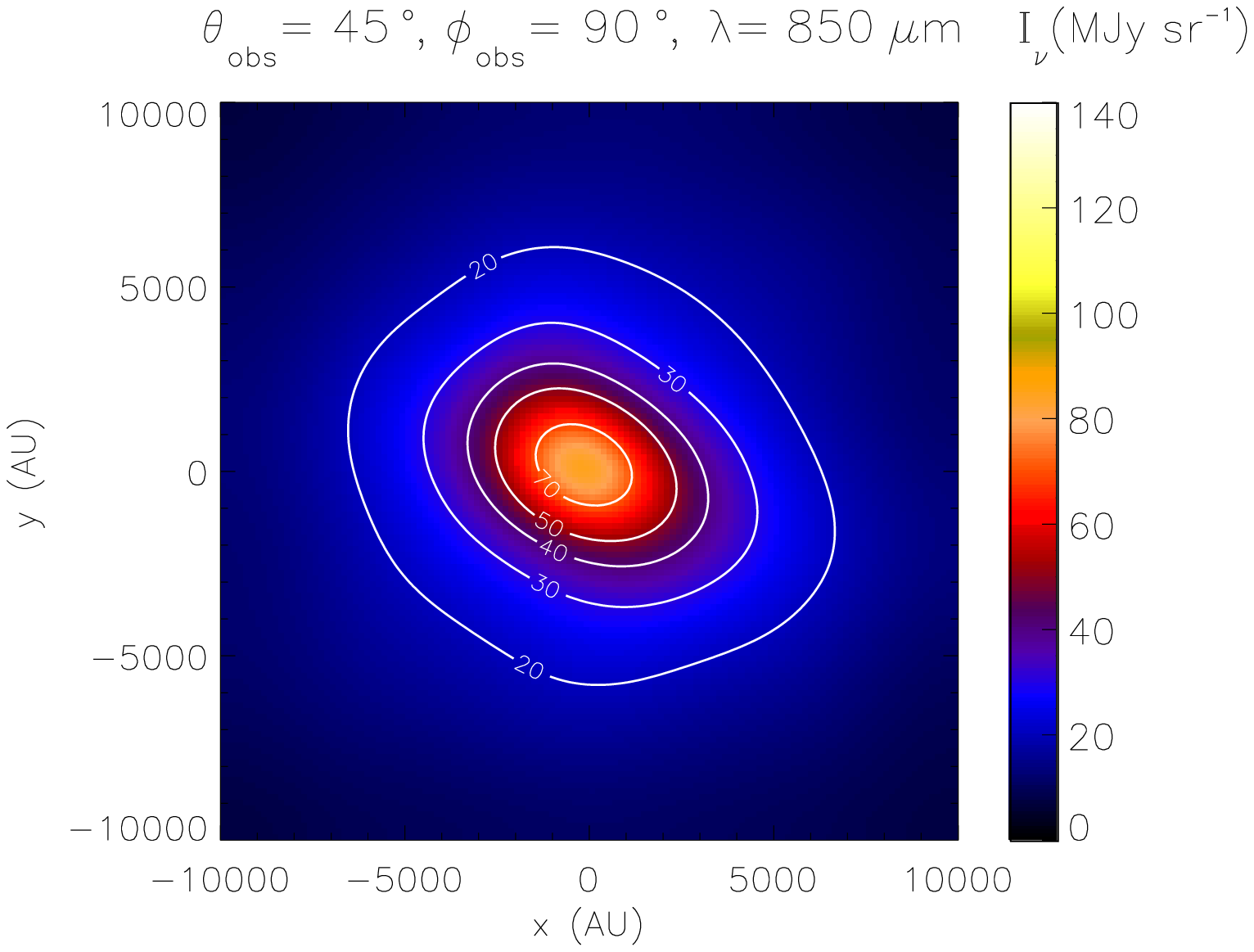}\hspace{-.3cm}
\includegraphics[width=6.65cm]{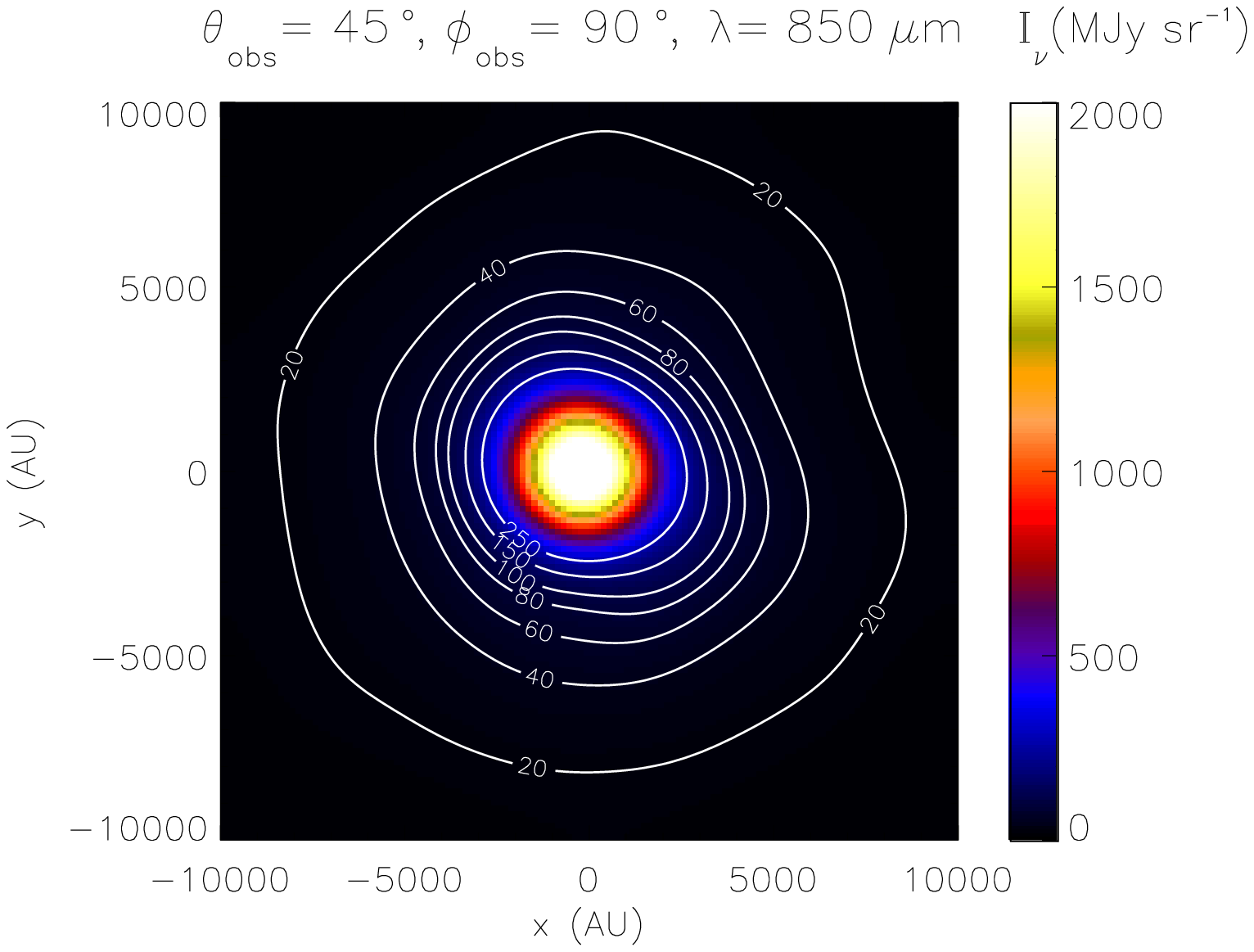}\hspace{-.3cm}
\includegraphics[width=6.65cm]{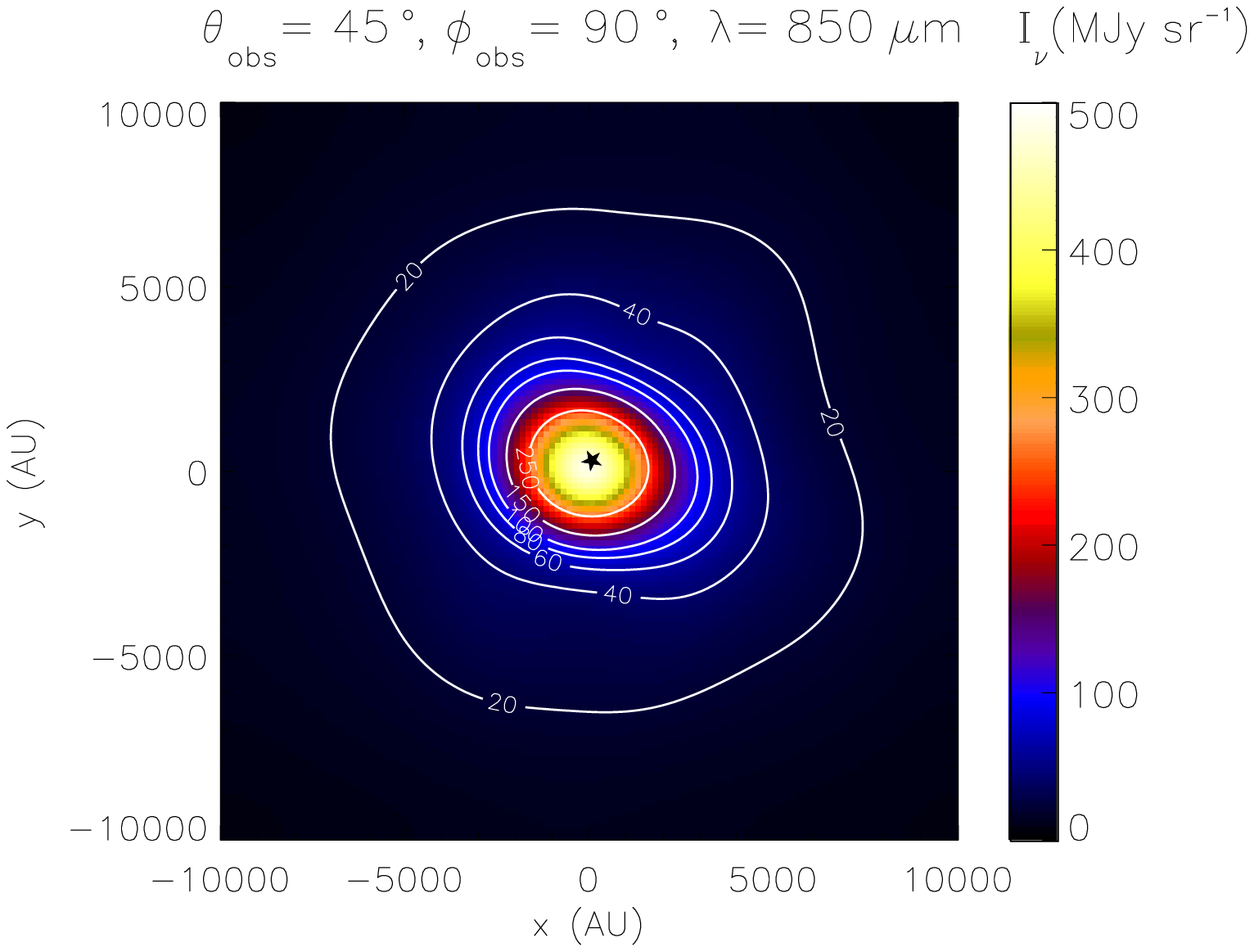}}
\caption{As Fig.~\ref{fig.isomaps}, but after convolving with 
a Gaussian beam having ${\rm FWHM} =  2300\,{\rm AU}$, (or equivalently 
$15\arcsec$ for a core at $150\,{\rm pc}$). Cores with protostars (middle 
frame, {\texttt t3};$\;$ right-hand frame, {\texttt t5}) appear more circular, 
but otherwise similar to the prestellar core (left-hand frame, {\texttt t1}).}
\label{fig.isomaps.conv}
\end{figure*}

\subsection{Isophotal maps of prestellar cores and young protostars}

The shapes of prestellar cores and Class 0 objects depend on viewing angle. 
In Fig.~\ref{fig.isomaps} we present isophotal maps at $850\,\micron$, for 
3 different time-frames ({\texttt t1}, {\texttt t3}, {\texttt t5}), and from 
3 different viewing angles. At $850\,\micron$ the core is optically thin, 
and the temperature does not vary much ($\sim 10\;{\rm to}\,20\,{\rm K}$, 
apart from the region very close to the protostar in time-frames {\texttt 
t3} and {\texttt t5}), so the maps are effectively column density maps. 
Class 0 objects are more centrally condensed than prestellar cores. They 
also show more structure, due to bipolar outflows, which clear low-density 
cavities (see images at $\theta = 90\degr$ and $\phi=90\degr$ on 
Fig.~\ref{fig.isomaps}).

In Fig.~\ref{fig.isomaps.conv} we present a selection of isophotal maps after 
convolving them with a Gaussian beam having ${\rm FWHM} = 2,300\,{\rm AU}$. 
Assuming the cloud is 150~pc away, this corresponds to an angular resolution 
of $15\arcsec$, which is similar to the beam size of current submm and mm 
telescopes (e.g. SCUBA, IRAM). On these maps, Class 0 objects look very 
similar to prestellar cores, apart from the fact that they tend to appear 
more circular and featureless (cf. Fig.~\ref{fig.isomaps}).

The reason for this is that the emission from a core that contains a protostar is 
concentrated in the central few hundred AU, and so, when it is convolved with 
a $2,300\,{\rm AU}$ beam, it produces an image rather like the beam, i.e. 
round and smooth. In contrast, the emission from a prestellar cores has structure 
on scales of several thousand AU, much of which survives convolution with a 
$2,300\,{\rm AU}$ beam. Thus we should expect cores that contain protostars to 
appear rounder than prestellar cores, and indeed this is what is observed 
(Goodwin et al. 2002).

\subsection{Individual time-frames}

{\it Time-frame} {\texttt t0}, {\it precollapse prestellar core.} The core is
heated exclusively by the ambient radiation field, and so the outer
parts of the cloud are warmer ($\sim\!20\,{\rm K}$) than the interior. The 
temperature falls to $\sim 8\,{\rm K}$ in the central $\sim 5000\,{\rm AU}$ 
of the cloud, where the density is high, $\sim 10^6$~cm$^{-3}$ (cf. Evans 
et al. 2001, Zucconi et al. 2001, Stamatellos \& Whitworth 2003, 
Gon{\c c}alves et al. 2004, Stamatellos et al. 2004). The cloud emission 
peaks at $\sim 190\,\micron$ which implies $T_{_{\rm EFF}} \sim 13\,{\rm K}$. 
The SED is fairly typical for a prestellar core (e.g. Ward-Thompson et al. 
2002), except that $T_{_{\rm EFF}}$ is a little higher than for the majority 
of prestellar cores (where it is $\sim\!10\,{\rm K}$). This is because the SED 
plotted here includes the outer, warmer layers of the core, and this shifts 
the peak of the SED to shorter wavelengths.

{\it Time-frame} {\texttt t1}, {\it collapsing prestellar core.} The collapse 
of the core results in a flattened, disc-like region in the centre of the cloud. 
Heating is still provided only by the ambient radiation field. The 
temperature at the centre of cloud  is even lower ($\sim 5\,{\rm K}$) than 
in time-frame {\texttt t0}, because the density -- and hence the optical depth -- 
is even higher ($> 10^8\,{\rm cm}^{-3}$). The SED looks very similar to time-frame 
{\texttt t0}, even though the collapse has started and the cloud is denser 
and colder at its centre. Thus the SED does not distinguish between a precollapse 
prestellar core and a collapsing prestellar core. 

{\it Time-frame} {\texttt t2}, {\it Class 0 object.} A protostar has formed at 
the centre of the core, and started heating the core. The protostar has low mass 
($\sim 0.01\,{\rm M}_{\sun}$), but the accretion rate onto it is high 
($\dot{M}_{\star}\sim 5\times10^{-5}{\rm M}_{\sun}$/yr), and hence the 
accretion luminosity is also quite high, $L_{_\star} \sim 5.7\,{\rm L}_\odot$. 
The dust temperature increases very close to the protostar, but drops down below 
$20\,{\rm K}$ within a few $1000\,{\rm AU}$ from the protostar, and then rises 
towards the edge where the core is heated by the ambient radiation field (cf. 
Shirley et al. 2002). Thus, at this early stage, it is necessary to probe the 
inner $1000\,{\rm AU}$ to infer the presence of the protostar. The peak of the 
SED has moved to $\sim 100\,{\rm to}\,120\,\micron$ (depending on viewing angle), 
which implies an effective temperature $T_{_{\rm EFF}} \sim 21\;{\rm to}\,25\,
{\rm K}$. The dependence of the SED on viewing angle is due to the disc around 
the protostar.

{\it Time-frame} {\texttt t3}, {\it Class 0 object.} The bipolar jets from the 
protostar have created hourglass cavities, as can be seen on the density plots of 
Fig.~\ref{fig.dens.temp}. As a consequence, the accretion rate onto the 
protostar decreases to $\dot{M}_{\star}\sim 10^{-5} {\rm M}_{\sun}$/yr, but 
its mass has increased (to $\sim 0.20\, {\rm M}_{\sun}$), and so the luminosity 
injected into the cloud increases (to $L_{_\star} \sim 27.2\,{\rm L}_{\sun}$). 
As a result, the temperature is $>25\,{\rm K}$ within $1000\;{\rm to}\;1500\,
{\rm AU}$, and it does not fall below $20\,{\rm K}$ in the inner $5000\,{\rm AU}$. 
The SED peaks at $\sim 80\;{\rm to}\,100\,\micron$ (depending on viewing angle), 
which implies $T_{_{\rm EFF}} \sim 25\;{\rm to}\,31\,{\rm K}$.

{\it Time-frame} {\texttt t4}, {\it Class 0 object.} The total luminosity injected 
into the cloud is lower ($L_{_\star} \sim 12.3\,{\rm L}_{\sun}$) due to the reduced 
accretion rate onto the protostar ($\dot{M}_{\star} \sim 2\times10^{-6}\,
{\rm M}_{\sun}$/yr).  The temperature distribution is similar to the previous 
time-frame ({\it time-frame} {\texttt t2}), but slightly cooler due to the reduced 
luminosity. The SED is also similar, but there is a slightly stronger dependence on 
viewing angle.  The SED now peaks at $\sim 90\;{\rm to}\,135\,\micron$, implying 
$T_{_{\rm EFF}} \sim 18\,{\rm to}\,28\,{\rm K}$.

{\it Time-frame} {\texttt t5}, {\it Class 0 object.} Asymmetries in the pattern 
of accretion onto the protostar have given it a small peculiar velocity, $\sim 
0.3\,{\rm km}\,{\rm s}^{-1}$, and by {\texttt t5} it is displaced from the dense 
central region by $\sim\!100$~AU. Consequently the accretion rate onto the 
protostar is reduced (to $\dot{M}_{\star} \sim 4 \times 10^{-7}\,{\rm M}_{\sun}$/yr) 
and with it the accretion luminosity (to $L_{_\star} \sim 2.5\,{\rm L}_{\sun}$). 
As illustrated in the temperature cross section parallel to the $x=0$ plane on 
Fig.~\ref{fig.dens.temp}, the upper half of the core is warmed to $> 14\,{\rm K}$ 
by the displaced protostar. In contrast, the lower half of the core is not affected 
by the protostar, and the temperature here does not rise above $\sim 10\,{\rm K}$, 
apart from in the outer layers of the core, which are still heated by the ambient 
radiation field. As a consequence of the reduced accretion luminosity, the peak of 
the SED moves back to longer wavelengths ($\sim\!150\,\micron$) and cooler effective 
temperature ($\sim\!17\,{\rm K}$). The displacement of the protostar also means that 
it is now visible at shorter wavelengths ($1\;{\rm to}\,50\,\micron$), from viewing 
angles close to the rotation axis.

\section{Summary}

We have simulated the collapse of a turbulent molecular core using SPH, 
and performed 3-dimensional Monte Carlo radiative transfer simulations at 
different stages of this collapse, to predict dust temperature fields, 
SEDs and isophotal maps. We focus on the initial stages of protostar formation, 
i.e. just before and just after the formation of a protostar, and derive 
criteria for distinguishing between genuine prestellar cores and cores that contain 
very young protostars. As pointed out by Masunaga et al. (1998), very young protostars 
are difficult to observe directly, because they are deeply embedded, and may 
be wrongly classified as prestellar cores. Hence it is important to consider 
ways in which their presence might be inferred indirectly. The main results 
of this study are as follows.

(i) A starless core is heated only by the ambient interstellar radiation field. 
Hence its outer regions are warm ($\sim~20~{\rm K}$) and the temperature drops 
towards the centre of the core (to $\sim7\,{\rm K}$). As the core collapses the 
central region becomes denser and colder ($\sim 5\,{\rm K}$), because the optical 
depth to the centre increases. Thus, prestellar cores that are about to form a 
protostar tend to be colder than precollapse prestellar cores.

(ii) When a protostar first forms in a core, its luminosity vaporizes the dust 
in its immediate vicinity, and heats the dust that survives up to the dust destruction 
temperature ($\sim 1500\,{\rm to}\,2000\,{\rm K}$). Initially, only the region 
within $\sim 1000\,{\rm AU}$ of the protostar is hotter than $20\,{\rm K}$, but
as the collapse proceeds the accretion luminosity of the protostar increases and the
presence of the protostar affects the entire core.  At this point, the heating of 
the core is mainly due to the accretion luminosity of the 
protostar, and the effect of the external radiation field is secondary.

(iii) The effective temperature of the SED follows the variation of the accretion 
luminosity, which initially increases due to the increase of the protostar mass 
but then decreases due to the decline in the mass accretion rate onto the protostar. 
Thus the SED of a prestellar core peaks at $\sim 190\,\micron$, corresponding to 
an effective temperature 
$T_{_{\rm EFF}} \sim  13\,{\rm K}$, but after a protostar has formed in the core, the 
peak of the SED moves to shorter wavelengths ($\sim 80\,\micron$, corresponding to 
$T_{_{\rm EFF}} \sim  31\,{\rm K}$) as the accretion luminosity increases, and then back 
to longer wavelengths ($\sim 130\,\micron$, corresponding to $T_{_{\rm EFF}} \sim 
19\,{\rm K}$) as the accretion luminosity decreases.

(iv) The peak of the SED of a Class 0 object depends only weakly on viewing angle. 
In general the peak is at longer wavelength when the Class 0 object is viewed in the 
plane of the surrounding accretion disc. It varies by at most $\pm 25\,\micron$, and 
is always $\leq\,150\,\micron$. Thus, it should be straightforward to distinguish 
Class 0 objects from prestellar cores, whose SEDs peak at $\sim 190\,\micron$.

(v) In our simulations, a newly-formed protostar cannot be observed in the NIR 
because it is too deeply embedded. This result reflects the high densities which 
our model predicts in the immediate surroundings of a newly-formed protostar.  
Additionally, it depends on the dust opacity there. Thus, the presence or absence 
of NIR emission from very young embedded protostars (confirmed by the presence 
of molecular outflows or compact cm radio emission), could be used to constrain the model.

(vi) At submm and mm wavelengths, isophotal maps of cores that contain protostars 
(Class 0 objects) tend to have more structure than maps of prestellar cores, but 
their emission also tends to be more strongly concentrated towards their centre. 
Consequently, after convolving with the beams of current telescopes, most of this 
central structure is lost and cores that contain protostars appear almost circular. 
In contrast, prestellar cores tend to have more extended emission and tend to 
maintain their shape after convolution.

Based on the above results we propose two criteria for identifying cores 
which -- despite appearing to be prestellar -- may harbour very young 
protostars:

(a) They are warm ($T>15\,{\rm K}$) as indicated by the peak of the SED of the 
core ($\lambda_{\rm peak} < 170\,\micron$). This criterion requires that the peak 
of the SED can be measured to an accuracy of $\sim 30\,\micron$, which should be 
feasible with observations in the far-IR from ISO (e.g. Ward-Thompson et al. 2002) 
and the upcoming Herschel mission (Andr\'e 2002), and in the mm/submm region from 
SCUBA (Kirk et al. 2005) and IRAM (Motte et al. 1998). We have presumed that the 
core is heated by the average interstellar radiation field, and the criterion 
would have to be modified for a core irradiated by a stronger or weaker radiation 
field.

(b) Their submm/mm isophotal maps are circular, at least in the central 
$2000\,{\rm to}\,4000\,{\rm AU}$. 

Using the above criteria, it is relatively easy to identify cores that may 
contain young protostars. These warm, circular cores can then be probed by 
other means (e.g. deep radio observations with the VLA, or sensitive NIR 
observations with Spitzer), to ascertain whether they in fact contain very 
young protostars. 
 
\begin{acknowledgements}

We thank P.~Andr\'e for providing the revised version of the Black (1994)
ISRF, that accounts for the PAH emission. We also thank D. Ward-Thompson
for useful discussions and suggestions. This work was partly supported
from the EC Research Training Network ``The Formation and Evolution of
Young Stellar Clusters'' (HPRN-CT-2000-00155), and partly by PPARC grant 
PPA/G/O/2002/00497.

\end{acknowledgements}

\end{document}